\documentclass[12pt]{iopart}
\usepackage{iopams} 
\usepackage{graphicx} 

\begin{document}

\title[Classification of spin and multipolar squeezing]{Classification of spin and multipolar squeezing}

\author{Emi Yukawa and Kae Nemoto}

\address{ National Institute of Informatics, 2-1-2, Hitotsubashi, Chiyoda-ku, Tokyo 101-8430, Japan} 
\ead{yukawa@nii.ac.jp}
\vspace{10pt}
\begin{indented}
\item[]May 2015
\end{indented}

\begin{abstract}
We investigate various types of squeezing in a collective su($2J+1$) system consisting of spin-$J$ particles ($J>1/2$).  
We show that the squeezing in the collective su($2J+1$) system can be classified into unitary equivalence classes, each of which 
is characterized by a set of squeezed and anti-squeezed observables forming an su($2$) subalgebra in the su($2J+1$) algebra. 
The dimensionality of the unitary equivalence class is fundamentally related to its squeezing limit. 
We also demonstrate the classification of the squeezing among the spin and multipolar observables in a collective su($4$) system. 
\end{abstract}


\vspace{2pc}

\submitto{\JPA}
%
%
%

\section{Introduction} 
Many quantum information protocols involve nonclassical states to achieve their quantum advantages.  For instance, quantum high precision measurements achieve sensitivities beyond the standard quantum limit by utilising nonclassical states.  The standard quantum limit is given by a coherent state, which satisfies the minimum uncertainty relation where quantum fluctuations are equally shared by any two quadrature amplitudes.  One way to break this limit is to squeeze a coherent state~\cite{Yuen}.  A squeezed state exhibits quantum fluctuations below the standard quantum limit in one quadrature at the sacrifice of larger quantum fluctuations in the other, which is directly applicable to achieve high precision measurements. To apply squeezed states to high precision measurements, it is important that squeezing can be achieved relatively easily.  Fortunately, squeezing can be achieved via quadratic Hamiltonian, and hence it does not require higher-order optical nonlinearity such as Kerr effect~\cite{Slusher,Wu}.  Both squeezing and the quantum advantages by squeezing in the high precision measurements have been demonstrated in experiments~\cite{Aasi, Taylor}. 

The idea of squeezing has been extended to spin systems~\cite{Kitagawa,Wineland1,Wineland2}.  An ensemble of spins can be considered as a collective spin when it satisfies the symmetry under particle permutations.  An ensemble of spin-$1/2$ systems can be treated as an su($2$) system in a high dimension dependent on the number of spins in the ensemble.  The coherent states of the su($2$) system can be defined as the orbit of the SU($2$) group action on a reference state~\cite{Arecchi}.  Usually, we take the lowest weight state as the reference state, analogous to the vacuum state in the optical coherent states.  Squeezing can then be introduced on the coherent state of the collective su($2$) system.  The spin squeezed states show quantum fluctuations below the standard quantum limit in one degree of freedom, similar to the optical squeezed states.    
In the spin-$1$ case~\cite{Mustecaplioglu,Yukawa,Colangelo}, the Hilbert space of a spin-$1$ particle can be spanned by three orthnormal states, and we can consider eight independent observables on the Hilbert space.  They correspond to eight generators of the su($3$) algebra, and hence the collective system inherits the su($3$) structure.  The dimensionality of the su($3$) collective system can be determined by the number of the spin-$1$ particles.  
Similarly, if the ensemble is of spin-$J$ systems, the collective spin can be treated as an su($2J+1$) system, where its dimension is determined by the number of the ensemble.  This extension is relevant to current experiments of squeezing on spin ensembles; for instance, squeezing in a spin-$7/2$ atomic gas~\cite{Auccaise} and in spin-1 Bose-Einstein condensates have been observed~\cite{Hald,Sewell,Hamley}.  In view of current and near future experimental developments, it is important to characterize the rich structure of squeezing in the collective su($2J+1$) systems and to systematically classify them based on unitary equivalent classes.

Among the collective su($2J+1$) systems, the su($2$) collective system is simple enough so that squeezing can be understood in comparison with optical squeezing. The representation space based on the SU($2$) coherent states is a sphere, i.e. the Bloch sphere.  Squeezing can be tracked on this two-dimensional space.  Though it is compact,  as the dimensionality of the Bloch sphere is the same as that of the phase space based on the optical coherent states, there are similarities between the SU($2$) squeezed states and the optical squeezed states.  When we extend the former to an ensemble of spin-$J$ systems ($J>1/2$), the structure of squeezing is no longer so simple.  As the collective su($2J+1$) system has $(2J+1)^2-1$ independent observables, there are a number of possible realizations of squeezed states.  In the case of $J=1$, there are the eight independent observables, which can be represented by three spin-vector components and five quadrupolar-tensor components, and squeezing can be implemented in terms of the su($2$) subalgebra among these eight observables.  Then, squeezing can be classified into two classes with the different squeezing limits.

In this paper, we generalize the classification to collective su($2J+1$) systems, where the squeezing can be characterized by 
$(2J+1)^2-1$ linearly independent observables.  Following the classification in Ref.~\cite{Yukawa}, we classify squeezing based on the unitary equivalence classes, whose definition is given in Sec. 2.2.  We also derive the structure factor of the su($2$) generators to characterise each class and obtain the squeezing limits via the one-axis twisting interaction. 

This paper is organized as follows. In Sec. II, we generalize the classification to the collective su($2J+1$) systems to show that the squeezing can be classified into the unitary equivalence classes of ($2J+1$)-dimensional representations of the su($2$) subalgebras.  In Sec. III, we derive quantum fluctuations for squeezed states of collective su($2J+1$) systems with one-axis twisting and show their squeezing limits.  In Sec. IV we apply our classification to a collective su($4$) system to illustrate the unitary equivalence classes of the squeezing and their squeezing limits, and summarize the main results in Sec. V. 
Throughout this paper, a scalar, a vector, and a matrix are respectively represented by a normal letter, a bold letter, and a normal letter with a tilde, as in $A$, $\bi{A}$, and $\tilde{A}$. The operator is denoted by a letter with a caret as in $\hat{A}$. 

\section{Classification of squeezing in collective su(2J+1) systems} 
\subsection{Observables of the su(2J+1) systems} 
Let us identify the linearly independent observables whose quantum fluctuations can be controlled via squeezing. 
Suppose there is a collective su($2J+1$) system consisting of $N$ spin-$J$ particles. 
The particles can be fermions as well as bosons when the spatial degrees of freedom of each fermion are frozen 
and the spin degrees of freedom are separable from the spatial degrees of freedom as in ultracold fermions trapped in an optical lattice~\cite{Martin} or 
magnetic impurities in a crystal~\cite{Tyryshkin}. 
We consider a squeezed spin state (SSS) which is generated from a coherent spin state (CSS) via a nonlinear interaction such as the one-axis twisting or the two-axis 
counter twisting~\cite{Kitagawa}. 

In a CSS, all particles are in the same single-spin state~\cite{Kitagawa}. 
A single-spin state can be expanded in terms of the rank-$d$ multipoles ($d\in \mathbb{N}$) and it can be described by the spherical harmonics of degree $d$. 
In the case of a spin-$1/2$ particle, the three components of the dipole, i.e., the spin vector, are linearly independent and generate the su($2$) algebra.  
In the case of a spin-$J$ particle, the $2d+1$ components of the rank-$d$ multipoles ($1 \leq d \leq 2J$) are linearly independent of each other, while the multipoles 
of the rank higher than $2J$ can be expressed in terms of the lower-rank multipoles and the identity. 
Thus, the spin and multipoles up to the rank of $2J$, which are comprised of $(2J+1)^2-1=4J(J+1)$ observables in total, completely characterize a single spin-$J$ state; hence they 
can be chosen as the generators of the su($2J+1$) algebra. 
We define the second-quantized forms of the single spin and multipolar observables as  
\begin{equation} 
	{\hat{\lambda}}_{{\bi n}_j;J,k} = \sum_{m,n=1}^{2J+1} ({\tilde{\lambda}}_{J,k})_{mn} {\hat{c}}^{\dagger}_{{\bi n}_j;J,m} {\hat{c}}_{{\bi n}_j;J,n}, 
	\label{eq:single-gen}
\end{equation} 
where $({\tilde{\lambda}}_{J,k})_{mn}$ represents the $mn$-entry of the $k$-th spin or multipolar matrix ${\tilde{\lambda}}_{J,k}$ of a single spin-$J$ 
particle, and ${\hat{c}}_{{\bi n}_j;J,m}$ ($ {\hat{c}}_{{\bi n}_j;J,m}^{\dagger}$) denotes the spin-$J$ bosonic or fermionic 
annihilation (creation) operator of the spatial mode ${\bi n}_j$ and the magnetic sublevel $m_z=J+1-m$. 
Here, we define ${\hat{\lambda}}_{{\bi n}_j;J,k}$ in Eq.~(\ref{eq:single-gen}) so that the first three observables are given by the Cartesian components of the 
spin vector, the next five are given by the five independent components of the quadrupolar tensor, the next seven are the seven independent components of 
the octupolar tensor~\cite{Shiina}, and so on. 
We also note that the matrices ${\tilde{\lambda}}_{J,k}$ are normalized so that their trace norms satisfy
\begin{equation}
	||{\tilde{\lambda}}_{J,k}||_{\mathrm{trace}}^2 = \sum_{m_z=-J}^J m_z^2 = \frac{1}{3} J(J+1)(2J+1). \label{eq:norm}
\end{equation} 

A CSS can be completely described by the collective observables of the single spin and multipolar observables given in Eq.~(\ref{eq:single-gen}). 
Squeezing can redistribute quantum fluctuations in these collective observables. 
The second-quantized forms of the collective observables ${\hat{\Lambda}}_{J,k}$ can be expressed as 
\begin{equation} 
	{\hat{\Lambda}}_{J,k} = \sum_{j=1}^N {\hat{\lambda}}_{{\bi n}_j;J,k}. \label{eq:coll-gen}
\end{equation} 
The observables ${\hat{\Lambda}}_{J,k}$ in Eq.~(\ref{eq:coll-gen}) satisfy the same commutation relations as 
${\tilde{\lambda}}_{J,k}$ in Eq.~(\ref{eq:single-gen}). 
This implies that they also generate the su($2J+1$) algebra and the matrices $\{ {\tilde{\lambda}}_{J,k} \}$ can be regarded as the irreducible representation 
of $\{ {\hat{\Lambda}}_{J,k} \}$ in the basis of $\{ | J, m_z \rangle \}$, which represents the basis of the single-spin magnetic sublevels 
with respect to the quantization axis along the $z$ axis. 
Thus, a collective observable ${\hat{O}}_J$ of the collective su($2J+1$) system can be expressed by a $(2J+1)$-dimensional matrix representation ${\tilde{O}}_J$ 
in the representation space of $V(\{ |J, m_z \rangle \})$ as follows:  
\begin{equation} 
	{\tilde{O}}_J = \sum_{k=1}^{4J(J+1)} v_{J,k} {\tilde{\lambda}}_{J,k}, \label{eq:coll-obs}
\end{equation}
where the real coefficients $v_{J,k}$ satisfy $\sum_{k=1}^{4J(J+1)} v_{J,k}^2 = 1$. 

\subsection{Classification based on unitary equivalence classes}
We consider squeezing among three observables $\{ {\hat{O}}_{J,k} \}$ ($k=1,2,3$) of the collective su($2J+1$) system, which form an 
su($2$) subalgebra of the su($2J+1$) algebra and satisfy the commutation relations given by 
\begin{equation} 
	[{\hat{O}}_{J,3}, {\hat{O}}_{J,\pm}] = \pm f {\hat{O}}_{J,3}, \label{eq:sub-su2}
\end{equation} 
where ${\hat{O}}_{J,\pm}\equiv {\hat{O}}_{J,1} \pm i {\hat{O}}_{J,2}$ and $f>0$ represents the magnitude of the structure constant. 
Note that $f$ in Eq.~(\ref{eq:sub-su2}) is not always $f=1$, since $\pm f$ are equivalent to the structure factors of the su($2J+1$) algebra. 

The squeezing among an su($2$) subalgebra $\{ {\hat{O}}_{J,k} \}$ can be classified based on the unitary equivalence class. 
The unitary equivalence class of the squeezing among $\{ {\hat{O}}_{J,k} \}$ can be determined by the $(2J+1)$-dimensional matrix representation of 
$\{ {\hat{O}}_{J,k} \}$ in the space of $V(\{ |J,m_z \rangle \})$ spanned by the basis $\{ |J,m_z \rangle \}$. 
The unitary equivalence class is defined as follows: Suppose $\{ {\tilde{X}}_k \}$ and $\{ {\tilde{X}}^{\prime}_k \}$ are the $n$-dimensional matrix 
representations of the semi-simple Lie algebra. 
Then, the representations $\{ {\tilde{X}}_k \}$ and $\{ {\tilde{X}}^{\prime}_k \}$ belong to the same unitary equivalence class, if there exists an SU($n$) transformation 
matrix $\tilde{U}$ such that $\tilde{U} {\tilde{X}}_k {\tilde{U}}^{\dagger} = {\tilde{X}}^{\prime}_k$ for $\forall k$. 

In our case, $\{ {\tilde{O}}_{J,k} \}$ is the $(2J+1)$-dimensional matrix representation of the generators of the su($2$) algebra, which is semi-simple; hence 
$\{ {\tilde{O}}_{J,k} \}$ should be completely reducible. 
The matrix representation $\{ {\tilde{O}}_{J,k} \}$ and its representation space $V(\{ |J,m_z \rangle \})$ can be decomposed into the direct sum of the 
lower dimensional irreducible representations of the su($2$) generators and their representation spaces, respectively.  
Suppose the dimension of the $l$-th irreducible representation is $2J_l+1$. 
Then, there exists an orthonormal basis set $\{ |J_l, m_l {\rangle}_l \}$ 
($m_l = -J_l,\cdots ,J_l$) such that the $l$-th irreducible representation of the su($2$) algebra is given by the spin matrices 
$\{ {\tilde{\lambda}}_{J_l,1}, {\tilde{\lambda}}_{J_l,2}, {\tilde{\lambda}}_{J_l,3} \}$ for a spin-$J_l$ particle (c.f. Eq.~(\ref{eq:single-gen})). 
The state $|J_l, m_l {\rangle}_l$ can be expressed as a linear combination of $|J, m_z \rangle$ ($m_z = -J, \cdots, J$), and $\{ |J_l, m_l{\rangle}_l \}$ 
and $\{ |J_{l^{\prime}}, m_{l^{\prime}} {\rangle}_{l^{\prime}} \}$ ($l \neq l^{\prime}$) are orthogonal to each other. 
Then, the completely reducible representation of $\{ {\tilde{O}}_{J,k} \}$ can be expressed as
\begin{equation} 
	{\tilde{O}}_{J,k} = f \bigoplus_{l=1}^{r} {\tilde{\lambda}}_{J_l,k}, \ 
	V(\{ |J,m_z \rangle \}) = \bigoplus_{l=1}^{r} V(\{ |J_l,m_l {\rangle}_l \}). \label{eq:direct-sum}
\end{equation} 
In Eq.~(\ref{eq:direct-sum}), $r$ expresses the number of the irreducible representations and the ``subspins'' $J_l$ satisfy $\sum_{l=1}^{r} (2J_l+1) = 2J+1$. 
The structure constant $f$ of $\{ {\tilde{O}}_{J,k} \}$ defined in Eq.~(\ref{eq:sub-su2}) is given by  
\begin{equation}
	f = \sqrt{\frac{J(J+1)(2J+1)}{\sum_{l=1}^r J_l (J_l+1) (2J_l+1)}}, \label{eq:f}
\end{equation}  
which can be derived from the irreducibility of $\{ {\tilde{\lambda}}_{J_l,k} \}$ and the normalization condition in Eq.~(\ref{eq:norm}). 
Here, we note that $\{ {\tilde{\lambda}}_{J_l,k} \}$ and $V(\{ |J_l,m_l {\rangle}_l \})$ are arranged so that $J_l$ satisfies 
\begin{equation} 
	0 \leq J_r \leq \cdots \leq J_2 \leq J_1 \leq J,  \label{eq:order}
\end{equation} 
and we define $\{ {\tilde{\lambda}}_{J_l=0,k} \} = \{ 0, 0, 0 \}$. 

If two sets of the generators of the su($2$) subalgebras, $\{ {\hat{O}}_{J,k} \}$ and $\{ {\hat{O}}^{\prime}_{J,k} \}$, belong to the same unitary 
equivalence class, $\{ {\hat{O}}^{\prime}_{J,k} \}$ and the representation space $V(\{ |J, m_z \rangle \})$ can be decomposed into 
\begin{equation} 
	{\tilde{O}}^{\prime}_{J,k} = f^{\prime} \bigoplus_{l=1}^{r^{\prime}} {\tilde{\lambda}}_{J_l^{\prime},k}, \ 
	V(\{ |J, m_z \rangle \}) = \bigoplus_{l=1}^{r^{\prime}} V(\{ |J_l^{\prime},  m_l^{\prime} {\rangle}_l \}), 
\end{equation} 
where $f^{\prime} = [J(J+1)(2J+1)/\sum_{l=1}^{r^{\prime}}J_l^{\prime} (J_l^{\prime}+1)(2J_l^{\prime}+1)]^{1/2}$, $m_l^{\prime} =-J_l^{\prime},\cdots ,J_l^{\prime}$, 
and 
\begin{equation} 
	r=r^{\prime} \ \land \ \forall l, J_l = J_l^{\prime}. 
	\label{eq:eqn-class}
\end{equation} 
Equation~(\ref{eq:eqn-class}) implies that the structure constants $f$ and $f^{\prime}$ are equal. 
If two sets of the generators of the su($2$) subalgebras, $\{ {\tilde{O}}_{J,k} \}$ and $\{ {\tilde{O}}^{\prime}_{J,k} \}$, do not belong to the same unitary 
equivalence class, Eq.~(\ref{eq:eqn-class}) does not hold, since a unitary matrix transforms the basis but it cannot change $r$ and $J_l$. 
The unitary equivalence classes of the su($2J+1$) algebra can be systematically found via the Dynkin diagram of the su($2J+1$) algebra as explained in 
\ref{subsec:Dynkin}. 

\subsection{\label{subsec:Dynkin}Dynkin diagram and unitary equivalence class} 
\begin{figure}[t]
\begin{center}
\includegraphics[width=15cm,clip]{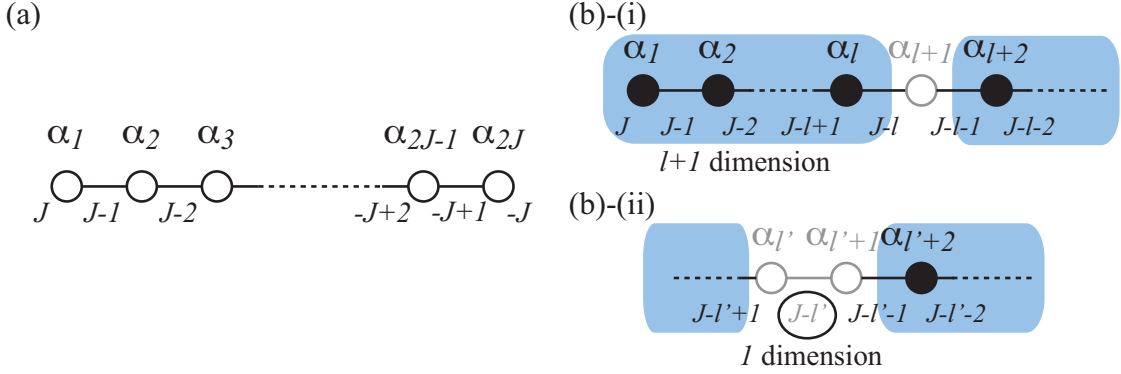}
\caption{(Color Online) (a) Dynikin diagram of the su($2J+1$) algebra. The simple root ${\balpha}_k$ expresses the transition from $m_z=J-k$ to $m_z=J-k+1$. 
(b) Correspondences between the connected and disconnected simple roots and the lower dimensional irreducible representations of the su($2$) generators. 
The filled circles and the gray open circles represent the simple roots that are chosen and  not chosen, respectively. 
(i) If the chosen simple roots from ${\balpha}_1$ to ${\balpha}_l$ are connected, then they are substituted by the $(l+1)$-dimensional irreducible representation in the 
representation space of $V(\{ |J, m_z \rangle \})$ ($m_z = J, \cdots , J-l$). 
(ii) If a magnetic sublevel $J-l^{\prime}$ is isolated from the connected simple roots, then it is substituted by the one-dimensional element, i.e., $0$.}
\label{fig:dynkin-sun}
\end{center}
\end{figure} 
The decomposition of the generators $\{ {\tilde{O}}_{J,k} \}$ of the su($2$) subalgebra in Eq.~(\ref{eq:direct-sum}) can be derived from the Dynkin 
diagram of the su($2J+1$) algebra. 
In the Dynkin diagram of the su($2J+1$) algebra, the $2J$ simple roots are connected as shown in Fig.~\ref{fig:dynkin-sun} (a). 
Here, the $k$-th vertex represents the $k$-th simple root ${\balpha}_k$ that corresponds to the raising matrix ${\tilde{A}}_{J,k}$ from the $k$-th sublevel to 
the $(k+1)$-th sublevel with respect to the quantization axis determined by the Cartan subalgebra. 
For the generators of the Cartan subalgebra, we choose the $z$ component of the spin vector ${\tilde{\lambda}}_{J,3}$ and the other $2J-1$ diagonal matrices. 
Then the quantization axis is given by the $z$ axis, which implies that ${\tilde{A}}_{J,k}$ raises the sublevel from $m_z = J-k$ to $m_z=J-k+1$ as follows: 
\begin{equation} 
	({\tilde{A}}_{J,k})_{mn} \equiv \sqrt{\frac{1}{3}J(J+1)(2J+1)} \ {\delta}_{J-k+1,m} {\delta}_{J-k,n}. \label{eq:simple-root-m} 
\end{equation} 
The matrix products of ${\tilde{A}}_{J,k}$ and their linear combinations reproduce the spin and multipolar observables ${\tilde{\lambda}}_{J,k}$.  

We can construct a complete irreducible representation by choosing $1\leq n\leq 2J$ vertices from the $2J$ vertices and substituting $l$-connected roots 
of ${\balpha}_{k}$, ${\balpha}_{k+1}$, $\cdots$, ${\balpha}_{k+l-1}$ ($l=1,\cdots ,2J$) by the $(l+1)$-dimensional irreducible representation $\{ {\tilde{\lambda}}_{J_l,k} \}$ 
($k=1,2,3$) of the su($2$) generators in the representation space of $V(\{ | J,m_z \rangle \})$ ($m_z = J-k+1, J-k, \cdots , J-k-l+1$) as shown 
in Fig.~\ref{fig:dynkin-sun} (b)-(i). 
If the magnetic sublevel of $m_z$ is not involved by the connected simple roots, then it is substituted by the one-dimensional element of $0$. 
This procedure is equivalent to the decomposition of $V(\{ | J,m_z \rangle \})$ into the subspaces in Eq.~(\ref{eq:direct-sum}) 
Arranging the irreducible representations so that their dimensions satisfy Eq.~(\ref{eq:order}), we can obtain the decomposition in Eq.~(\ref{eq:direct-sum}). 
Since the Dynkin diagram does not depend on the choice of the basis, any $(2J+1)$-dimensional matrix representation can be obtained by rotating 
one of the representations derived from the Dynkin diagram via an SU($2J+1$) unitary matrix. 

\section{Properties of squeezing determined by unitary equivalence classes} 
\subsection{Squeezing parameters}
The properties of the squeezing reflect the structure of the unitary equivalence class, i.e., the subspins and the initial coherent state. 
To confirm this, let us consider squeezing among an su($2$) subalgebra $\{ {\hat{O}}_{J,k} \}$, which can be decomposed into Eq.~(\ref{eq:direct-sum}) 
with the subspins $\{ J_l \}$. 
A CSS~\cite{Perelomov,Gilmore,Nemoto,Mathur} can be expressed in terms of two parameters 
$\theta \in [0,\pi]$ and $\phi \in [0,2\pi )$ as 
\begin{eqnarray}
	&|\theta ,\phi {\rangle}_{\mathrm{tot}} 
	\equiv {\left [ \bigoplus_{l=1}^r {\zeta}_l |\theta ,\phi {\rangle}_l \right ]}^{\otimes N} 
	\nonumber \\
	&= \sum_{n_1=0}^N \sum_{n_2=0}^{N-n_1} \cdots \sum_{n_{r-1}=0}^{N-n_1-\cdots - n_{r-2}} \sqrt{_NC_{n_1} \ _{N-n_1}C_{n_2} \
	\cdots \ _{N-n_1-\cdots - n_{r-2}}C_{n_{r-1}}} \nonumber \\
	&\times {\zeta}_1^{n_1} {\zeta}_2^{n_2} \cdots {\zeta}_{r-1}^{n_{r-1}} {\zeta}_r^{N-n_1-\cdots - n_{r-1}} \nonumber \\
	&\times \Bigl [  |\theta ,\phi {\rangle}_1^{\otimes n_1} \oplus |\theta ,\phi {\rangle}_2^{\otimes n_2} 
	\oplus \cdots \oplus |\theta ,\phi {\rangle}_{r-1}^{\otimes n_{r-1}} \oplus |\theta ,\phi {\rangle}_r^{\otimes N - n_1 - \cdots - n_{r-1}} \Bigr ], 
	 \label{eq:CSS-J}
\end{eqnarray} 
where $\sum_{l=1}^r |{\zeta}_l|^2 = 1$ and the single particle states $|\theta , \phi {\rangle}_l$ in Eq.~(\ref{eq:CSS-J}) for $J_l \neq 0$ and $J_l = 0$ are 
defined in terms of the basis $\{ |J_l, m_l {\rangle}_l \}$ as 
\begin{equation} 
	\forall J_l \neq 0, \ |\theta , \phi {\rangle}_l \equiv \exp {\left [-\frac{\theta}{2} ( e^{-i\phi } {\tilde{\lambda}}_{J_l,+} - e^{i\phi } {\tilde{\lambda}}_{J_l,-}) \right ]}  
	|J_l, J_l {\rangle}_l, \label{eq:one-J} 
\end{equation} 
with ${\tilde{\lambda}}_{J_l,\pm} \equiv {\tilde{\lambda}}_{J_l,1} \pm i {\tilde{\lambda}}_{J_l,2}$, 
and $|\theta ,\phi {\rangle}_l \equiv |J_l = 0, m_l = 0 {\rangle}_l$ ($J_l = 0$), respectively. 
The CSS $|\theta ,\phi {\rangle}_{\mathrm{tot}}$ in Eq.~(\ref{eq:CSS-J}) satisfies the minimum uncertainty relation 
\begin{equation} 
	\forall \nu \in [0,2\pi ), \ \langle ( \Delta O_{J,\nu} )^2 \rangle \langle ( \Delta O_{J,\nu +\frac{\pi}{2}} )^2 \rangle 
	= \frac{f^2}{4} \langle {\hat{O}}_{J, \perp} {\rangle}^2, \label{eq:MUR}
\end{equation} 
where $\langle \hat{X} \rangle$ represents the expectation value of an observable $\hat{X}$, the quantum fluctuation in $\hat{X}$ is defined 
as $\langle (\Delta X )^2 \rangle = \langle {\hat{X}}^2 \rangle - \langle \hat{X} {\rangle}^2$, and 
${\hat{O}}_{J,\nu}$ and ${\hat{O}}_{J, \perp}$ are given by 
\begin{eqnarray} 
	& {\hat{O}}_{J,\perp} \equiv {\hat{O}}_{J,1} \cos {\phi} \sin {\theta} + {\hat{O}}_{J,2} \sin {\phi} \sin {\theta} + {\hat{O}}_{J,3} \cos {\theta}, 
	\label{eq:O3} \\
	& {\hat{O}}_{J,\nu} \equiv {\hat{O}}_{J,1} ( \cos {\phi} \cos {\theta} \cos {\nu} - \sin {\phi} \sin {\nu} ) \nonumber \\ 
	&+ {\hat{O}}_{J,2} ( \sin {\phi} \cos {\theta} \cos {\nu} + \cos {\phi} \sin {\nu} ) 
	- {\hat{O}}_{J,3} \sin {\theta} \cos {\nu},  \label{eq:O1} 
\end{eqnarray} 
respectively. 
The expectation values in Eq.~(\ref{eq:MUR}) can be obtained via the Schwinger-boson approach described in \ref{a:0} as 
\begin{eqnarray} 
	& \langle {\hat{O}}_{J, \perp} \rangle = fN \sum_{l,J_l\neq 0} J_l |{\zeta}_l|^2, \label{eq:exp-CSS} \\
	& \forall \nu \in [0, 2\pi ), \ \langle ( \Delta O_{J,\nu} )^2 \rangle = \frac{f^2N}{2} \sum_{l,J_l\neq 0} J_l |{\zeta}_l|^2.  \label{eq:fluct-CSS} 
\end{eqnarray} 
Equations~(\ref{eq:MUR}), (\ref{eq:exp-CSS}) and (\ref{eq:fluct-CSS}) imply that the squeezing among $\{ {\hat{O}}_{J,k} \}$ can suppress 
$\langle ( \Delta O_{J,\nu} )^2 \rangle$ below the 
coherent-spin-state value of $\frac{f}{2} |\langle {\hat{O}}_{J, \perp} \rangle |$ at the expense of $\langle ( \Delta O_{J,\nu+\frac{\pi}{2}} )^2 \rangle$ 
enhanced above $\frac{f}{2} |\langle {\hat{O}}_{J, \perp} \rangle |$; hence, the squeezing can be characterized by the squeezing parameter $\xi$ defined as  
\begin{equation}
	{\xi}^2 = \left ( 2N \sum_{l,J_l\neq 0} J_l |{\zeta}_l|^2 \right ) \times \frac{\min_{\nu} 
	{\langle (\Delta O_{J, \nu} )^2 \rangle} }{\langle {\hat{O}}_{J, \perp} {\rangle}^2}, 
	\label{eq:sqparam}
\end{equation} 
where $\min_{\nu} {\langle (\Delta O_{J, \nu} )^2 \rangle}$ is the quantum fluctuations in Eq.~(\ref{eq:O1}) perpendicular to the $O_{J, \perp}$ plane and minimized with 
respect to the angle $\nu$ in Eq.~(\ref{eq:O1}). 
Equation~(\ref{eq:sqparam}) is equal to $1$ for the CSS in Eq.~(\ref{eq:CSS-J}) and it implies that a state giving ${\xi}^2 < 1$ is squeezed. 
We note that Eq.~(\ref{eq:sqparam}) is equivalent to the Wineland's squeezing parameter~\cite{Wineland1} when the coefficients $\{ |{\zeta}_l |^2 \}$ of the 
initial CSS in Eq.~(\ref{eq:CSS-J}) are given by $|{\zeta}_l|^2 = {\delta}_{l,l_0}$ with $l_0$ such that $J_{l_0} \neq 0$. 
The squeezing parameter $\xi$ in Eq.~(\ref{eq:sqparam}) is characterized by the subspins and the initial CSS, both of which reflect 
the structure of the unitary equivalence class of the spin and multipolar observables $\{ {\hat{O}}_{J,k} \}$ generating the su($2$) subalgebra. 

\subsection{Squeezed and anti-squeezed quantum fluctuations for one-axis twisting interactions} 
Let us calculate the squeezing parameter $\xi$ in Eq.~(\ref{eq:sqparam}) for an SSS generated via the one-axis twisting interaction~\cite{Kitagawa}. 
We consider the one-axis twisting interaction 
\begin{equation}
	{\hat{H}}_{\mathrm{OAT}} = \hbar \chi {\hat{O}}_{J,3}^2 \label{eq:OAT} 
\end{equation}
with the interaction energy $\chi$, which distribute the quantum fluctuations in the $O_{J,2}$-$O_{J,3}$ plane. 
A CSS of the $N$ spin-$J$ particles is given by 
$| \theta = \frac{\pi}{2}, \phi = 0 {\rangle}_{\mathrm{tot}}$ in Eq.~(\ref{eq:CSS-J}). 
Defining the rescaled evolution time $\mu \equiv 2 \chi f^2 t$, we can express the one-axis-twisted SSS 
$| {\Psi}_{\mathrm{OAT}} (J,N;\mu ) {\rangle}_{\mathrm{tot}}$ at $\mu $ as 
\begin{equation} 
	| {\Psi}_{\mathrm{OAT}} (J,N;\mu ) {\rangle}_{\mathrm{tot}} = \exp {\left [- \frac{i}{2f^2} {\hat{O}}_{J,3}^2 \mu\right ]} \ 
	|\theta = \frac{\pi}{2}, \phi = 0 {\rangle}_{\mathrm{tot}}. \label{eq:SSS-J} 
\end{equation} 
In this case, the observable ${\hat{O}}_{J,\perp}$ is given by ${\hat{O}}_{J,1}$ and its expectation value at time $\mu$ can be obtained in a manner similar to 
Eqs.~(\ref{eq:exp-CSS}) and (\ref{eq:fluct-CSS}) as 
\begin{eqnarray} 
	\langle  {\hat{O}}_{J, 1}  \rangle (\mu ) &\equiv \langle  {\Psi}_{\mathrm{OAT}} (J,N;\mu ) | {\hat{O}}_{J,1} | {\Psi}_{\mathrm{OAT}} (J,N;\mu ) {\rangle}_{\mathrm{tot}} \nonumber \\
	&= f N \sum_{l:J_l\neq 0} J_l |{\zeta}_l|^2 {\cos}^{2J_l-1} \frac{\mu}{2} \ {\left [1 - |{\zeta}_l|^2 \left ( 1 - {\cos}^{2J_l} \frac{\mu}{2} \right ) \right ] }^{N-1},  
	\label{eq:perp} 
\end{eqnarray} 
as detailed in \ref{a:0}. 
The quantum fluctuations in the plane perpendicular to ${\hat{O}}_{J,\perp}$ can be simplified as a function of $\nu$ as 
\begin{equation}
	\langle (\Delta O_{J, \nu} )^2 \rangle (\mu ) = \frac{f^2N}{2} 
	\sum_{l:J_l\neq 0} J_l|{\zeta}_l|^2 \left [ 1 +  A_l (1+ \cos {2\nu} ) - B_l \sin {2\nu} \right ]. 
	\label{eq:fluct-3} 
\end{equation} 
Here, $A_l$ and $B_l$ are defined as 
\begin{eqnarray}
	A_l & \equiv \frac{J_l }{2} (N-1) |{\zeta}_l|^2 \left \{ 1 - {\cos}^{2(2J_l-1)} \mu \ [1-|{\zeta}_l|^2 (1-{\cos}^{2J_l} \mu) ]^{N-2}
	\right \} \nonumber \\
	&+ \frac{1}{2} \left ( J_l - \frac{1}{2} \right ) \{ 1 - {\cos}^{2(J_l-1)} \mu \ [1-|{\zeta}_l|^2 (1-{\cos}^{2J_l} \mu) ]^{N-1} \}, \\
	B_l & \equiv 2 \Biggl \{ J_l (N-1) |{\zeta}_l|^2 {\cos}^{2J_l} \frac{\mu}{2} 
	+ \left ( J_l - \frac{1}{2} \right ) \left [1 - |{\zeta}_l|^2 \left (1 - {\cos}^{2J_l} \frac{\mu}{2} \right ) \right ] \Biggr \} \nonumber \\
	& \times \sin {\frac{\mu}{2}} \ {\cos}^{2(J_l-1)} \frac{\mu}{2} \ {\left [1 - |{\zeta}_l|^2 \left (1 - {\cos}^{2J_l} \frac{\mu}{2} \right ) \right ]}^{N-2}. \label{eq:fluct-4}
\end{eqnarray}  
Equation~(\ref{eq:fluct-3}) is periodic with respect to $\nu$, so there exist the minimum and the maximum, i.e., the squeezed and anti-squeezed quantum 
fluctuations, respectively. 
The squeezing parameter ${\xi}^2 (\mu =0) = 1$ for the initial CSS in Eq.~(\ref{eq:CSS-J}) and the spins are said to be squeezed when ${\xi}^2 (\mu )<1$. 

The squeezing limit in Eq.~(\ref{eq:sqparam}) can be analytically obtained in the limit of $\mu \ll 1$ and $N \gg 1$, when the subspins $\{ J_l \}$ in Eq.~(\ref{eq:direct-sum}) 
and the coefficients $\{ |{\zeta}_l|^2 \}$ of the initial coherent state in Eq.~(\ref{eq:CSS-J}) satisfy $|{\zeta}_l|^2 = {\delta}_{l,l_0}$ ($J_{l_0} \neq 0$). 
The quantum fluctuations in the $O_{J,2}$-$O_{J,3}$ plane in Eq.~(\ref{eq:fluct-3}) can be simplified as 
\begin{eqnarray} 
	\langle (\Delta O_{J_{l_0}, \nu} )^2 \rangle (\mu ) 
	&= \frac{f^2J_{l_0}N}{2} \Biggl \{ 1 + \frac{1}{2} \left (J_{l_0}N - \frac{1}{2} \right ) \nonumber \\
	& \times \Biggl [ ( 1 - {\cos}^{2(J_{l_0}N-1)} \mu ) (1 + \cos {2\nu} ) \nonumber \\
	&- 4 \sin {\frac{\mu}{2}} \ {\cos}^{2(J_{l_0}N-1)} \frac{\mu}{2} \ \sin {2\nu} \Biggr ] \Biggr \}, \label{eq:fluc-r1}
\end{eqnarray} 
and the expectation value perpendicular to the $O_{J,2}$-$O_{J,3}$ plane in Eq.~(\ref{eq:perp}) is 
\begin{equation} 
	\langle {\hat{O}}_{J, 1}  \rangle (\mu ) = fJ_{l_0}N {\cos}^{2JN-1} \frac{\mu}{2}. \label{eq:prep-r1}
\end{equation} 
Here, we assume that $\mu$ and $N$ satisfy $\alpha \equiv \frac{1}{2} J_{l_0}N \mu \gg 1$ and $\beta \equiv \frac{1}{4} J_{l_0}N {\mu}^2 \ll 1$. 
Then, substituting Eqs.~(\ref{eq:fluc-r1}) and (\ref{eq:prep-r1}) into Eq.~(\ref{eq:sqparam}), we obtain the squeezing parameter for $r=1$ up to the 
second order in $\beta$ as: 
\begin{equation} 
	{\xi}^2 (\mu ) \simeq \frac{1}{4{\alpha}^2} + \frac{2}{3} {\beta}^2 + \frac{\beta}{2{\alpha}^2} + \mathcal{O} (\max {\{ \frac{{\beta}^2}{\alpha}, {\beta}^3 \} }), 
	\label{eq:sqpram-r1}
\end{equation} 
where $\nu \simeq - \frac{1}{2} \arctan {\frac{1}{\alpha}} + \frac{\pi}{2}$. 
The minimum of Eq.~(\ref{eq:sqpram-r1}), i.e., the squeezing limit is achieved at $\mu = {\mu}_{\mathrm{min}} = (12)^{1/6} (J_{l_0}N)^{-2/3}$ are given by 
\begin{equation} 
	{\xi}_{\mathrm{min}}^2 \equiv {\xi}^2 ({\mu}_{\mathrm{min}}) \simeq \frac{1}{2} {\left ( \frac{3}{2J_{l_0}N} \right )}^{2/3} + \frac{1}{2J_{l_0}N} \propto {(J_{l_0}N)^{-2/3}}, 
	\label{eq:minsqparam-r1}
\end{equation} 
which implies that the squeezing limit monotonically decreases with increasing $J_{l_0}$. 

\section{Application to collective su($4$) systems} 
\subsection{Complete set of collective spin and multipolar observables} 
To examine the squeezing parameter in Eq.~(\ref{eq:sqparam}) for $r>1$, especially the $\{ |{\zeta}_l|^2 \}$-dependence of the squeezing limit, let us consider a collective 
su($4$) system consisting of $N$ spin-3/2 particles as an example. 
In this case, the observables that can completely characterize collective spin states are the spin vector, the quadrupolar tensor, and the 
octupolar tensor. 
The Cartesian components of the spin vector ${\hat{\lambda}}_{{\bi n}_j;J=\frac{3}{2},k}$ ($k = 1,2,3$) can be given by 
\begin{equation} 
	{\hat{\lambda}}_{{\bi n}_j;\frac{3}{2},k} = \sum_{j=1}^N \sum_{m,n=1}^4 ({\tilde{\lambda}}_{\frac{3}{2},k})_{mn} {\hat{c}}^{\dagger}_{{\bi n}_j;\frac{3}{2},m} 
	{\hat{c}}_{{\bi n}_j;\frac{3}{2},n}, \label{eq:single-j}
\end{equation} 
where ${\tilde{\lambda}}_{\frac{3}{2},k}$ represent the spin-$3/2$ matrices ${\tilde{J}}_{\mu}$ ($\mu = x,y,z$) given by Eq.~(\ref{eq:single-mj}). 
The matrix representations of the five independent components of the quadrupolar tensor and the seven independent components of the 
octupolar tensor~\cite{Shiina} can be respectively expressed in terms of ${\tilde{J}}_{\mu }$ as 
\begin{eqnarray} 
	({\tilde{Q}}_{\mu \nu})_{mn}= \frac{\sqrt{15}}{6}  ({\tilde{J}}_{\mu} {\tilde{J}}_{\nu} + {\tilde{J}}_{\nu} {\tilde{J}}_{\mu} )_{mn}, \label{eq:single-q} \\
	({\tilde{D}}_{xy})_{mn} = \frac{\sqrt{15}}{6} ({\tilde{J}}_{x}^2 - {\tilde{J}}_{y}^2 )_{mn}, \label{eq:single-d} \\
	(\tilde{Y})_{mn} = \frac{\sqrt{5}}{6} (-{\tilde{J}}_{x}^2 - {\tilde{J}}_{y}^2 + 2{\tilde{J}}_{z}^2 )_{mn}, \label{eq:single-y}
\end{eqnarray} 
where $(\mu ,\nu) = (x,y), (y,z), (z,x)$ in Eq.~(\ref{eq:single-q}), and 
\begin{eqnarray} 
	({\tilde{T}}^{\alpha}_{\mu})_{mn} = \frac{1}{3} (2 {\tilde{J}}_{\mu}^3 - \overline{{\tilde{J}}_{\mu} {\tilde{J}}_{\nu}^2} 
	- \overline{{\tilde{J}}_{\eta}^2 {\tilde{J}}_{\mu}})_{mn}, \label{eq:single-ta} \\
	({\tilde{T}}^{\beta}_{\mu})_{mn} = \frac{\sqrt{15}}{9} (\overline{{\tilde{J}}_{\mu} {\tilde{J}}_{\nu}^2} 
	- \overline{{\tilde{J}}_{\eta}^2 {\tilde{J}}_{\mu}})_{mn} \\
	({\tilde{T}}_{xyz})_{mn} = \frac{\sqrt{15}}{9} (\overline{{\tilde{J}}_{x}{\tilde{J}}_{y}{\tilde{J}}_{z}})_{mn}, \label{eq:single-txyz}
\end{eqnarray} 
where $(\mu ,\nu ,\eta) = (x,y,z)$, $(y,z,x)$, and $(z,x,y)$ and the overbars above the matrix products are defined as 
$\overline{\tilde{A} {\tilde{B}}^2} = \tilde{A} {\tilde{B}}^2 + \tilde{B} \tilde{A} \tilde{B} + {\tilde{B}}^2\tilde{A}$ and 
$\overline{\tilde{A} \tilde{B} \tilde{C}} = \tilde{A} \tilde{B} \tilde{C} + \tilde{B} \tilde{C} \tilde{A} + \tilde{C} \tilde{A} \tilde{B} + \tilde{B} \tilde{A} \tilde{C} 
+ \tilde{C} \tilde{B} \tilde{A} + \tilde{A} \tilde{C} \tilde{B}$ with respect to the matrices $\tilde{A}$, $\tilde{B}$, and $\tilde{C}$. 
Here we note that the matrix representations of the spin and multipolar observables in Eqs.~(\ref{eq:single-j})-(\ref{eq:single-txyz}) are normalized 
so that they satisfy the condition in Eq.~(\ref{eq:norm}). 
These fifteen spin and multipolar observables in Eqs.~(\ref{eq:single-j})-(\ref{eq:single-txyz}) together form the su($4$) Lie algebra. 
Then, the irreducible representations of the collective spin observables describing the symmetric spin state can respectively be 
given by the matrix representations of their single-spin counter parts in Eqs.~(\ref{eq:single-j})-(\ref{eq:single-txyz}), whose explicit 
expressions are given in Eqs.~(\ref{eq:single-mj})-(\ref{eq:single-mt}). 
We define the matrices $\{ {\tilde{\lambda}}_{\frac{3}{2},k} \} \equiv \{ {\tilde{J}}_{\mu}, {\tilde{Q}}_{\mu \nu}, {\tilde{D}}_{xy}, \tilde{Y}, {\tilde{T}}^{\alpha}_{\mu}, 
{\tilde{T}}^{\beta}_{\mu}, {\tilde{T}}_{xyz} \}$ ($k=1,\cdots ,15$) in the order of Eqs.~(\ref{eq:single-j})-(\ref{eq:single-txyz}). 
Then, the matrix representation of any observable can be expressed in terms of $\{ {\tilde{\lambda}}_{\frac{3}{2},k} \}$ ($k=1,\cdots ,15$) as in Eq.~(\ref{eq:coll-obs}). 

\subsection{Four types of squeezing} 
\begin{figure}[t]
\begin{center} 
\includegraphics[width=15cm,clip]{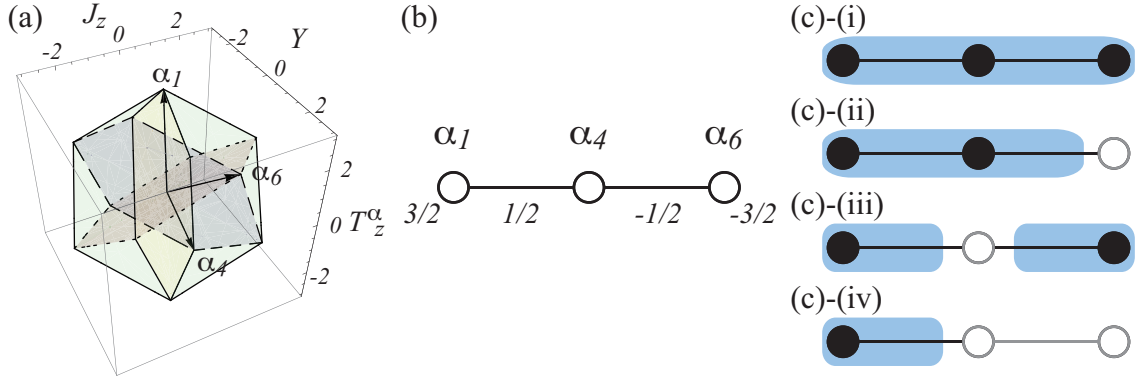}
\caption{(Color Online) (a) The root diagram of the su($4$) algebra, (b) the Dynkin diagram of the the su($4$) algebra, and (c) the four types of the 
unitary equivalence classes of the matrix representations of the su($2$) subalgebras. In (c), the chosen simple roots and the omitted simple roots are indicated 
by the filled black circles and the open grey circles, respectively. }
\label{fig:su4-roots}
\end{center}
\end{figure}

There exist four unitary equivalence classes of the su($2$) subalgebras in the su($4$) algebra, which can be found as explained in 
Sec.~\ref{subsec:Dynkin}. 
First, let us construct the Dynkin diagram and consider the relation between the simple roots and the spin and multipolar observables in 
Eqs.~(\ref{eq:single-j})-(\ref{eq:single-txyz}). 
In collective su($4$) systems, the Dynkin diagram has three simple roots ${\balpha}_1$, ${\balpha}_4$, ${\balpha}_6$ as shown in Fig.~\ref{fig:su4-roots} (b). 
Choosing the diagonal matrices ${\tilde{J}}_z = {\tilde{\lambda}}_{\frac{3}{2},3}$, $\tilde{Y} = {\tilde{\lambda}}_{\frac{3}{2},8}$, and 
${\tilde{T}}^{\alpha}_z = {\tilde{\lambda}}_{\frac{3}{2},11}$ as the generators of the Cartan subalgebra, we can express the matrices 
${\tilde{A}}_{\frac{3}{2},1}$, ${\tilde{A}}_{\frac{3}{2},4}$, and ${\tilde{A}}_{\frac{3}{2},6}$ corresponding to the simple roots as 
\begin{eqnarray} 
	{\tilde{A}}_{\frac{3}{2},1} 
	= \frac{\sqrt{15}}{10} {\tilde{J}}_+ + \frac{1}{2} {\tilde{Q}}_+ - \frac{\sqrt{15}}{20} {\tilde{T}}^{\alpha}_+ - \frac{1}{4} {\tilde{T}}^{\beta}_-,  \label{eq:a1} \\ 
	{\tilde{A}}_{\frac{3}{2},4} 
	= \frac{1}{\sqrt{5}} {\tilde{J}}_+ + \frac{3}{4\sqrt{5}} {\tilde{T}}^{\alpha}_+ + \frac{\sqrt{3}}{4} {\tilde{T}}^{\beta}_-, \label{eq:a4} \\
	{\tilde{A}}_{\frac{3}{2},6} 
	= \frac{\sqrt{15}}{10} {\tilde{J}}_+ - \frac{1}{2} {\tilde{Q}}_- - \frac{\sqrt{15}}{20} {\tilde{T}}^{\alpha}_+ - \frac{1}{4} {\tilde{T}}^{\beta}_-, \label{eq:a6} 
\end{eqnarray}  
where we define ${\tilde{J}}_{\pm} \equiv {\tilde{J}}_x \pm i{\tilde{J}}_y$, 
${\tilde{Q}}_{\pm} \equiv {\tilde{Q}}_{zx} \pm i{\tilde{Q}}_{yz}$,  
${\tilde{T}}^{\alpha}_{\pm} = {\tilde{T}}^{\alpha}_x \pm i{\tilde{T}}^{\alpha}_y$, and 
${\tilde{T}}^{\beta}_{\pm} = {\tilde{T}}^{\beta}_x \pm i{\tilde{T}}^{\beta}_y$, respectively. 
The derivation of Eqs.~(\ref{eq:a1})-(\ref{eq:a6}) are detailed in \ref{a:2}. 

Then, the four unitary equivalence classes of the su($2$) subalgebras can be found, that is, the types (i)-(iv) as illustrated in Figs.~\ref{fig:su4-roots} (c). 
The su($2$) subalgebra $\{ {\tilde{O}}_{\frac{3}{2},k} \}$ ($k=1,2,3$) of these four classes satisfy 
$[{\tilde{O}}_{\frac{3}{2},\pm}, {\tilde{O}}_{\frac{3}{2},3}]=\pm f{\tilde{O}}_{\frac{3}{2},\pm}$, where 
${\tilde{O}}_{\frac{3}{2},\pm} = {\tilde{O}}_{\frac{3}{2},1} \pm i{\tilde{O}}_{\frac{3}{2},2}$. 
Suppose the matrices $\{ {\tilde{O}}_{\frac{3}{2},k} \}$ have the block-diagonalized forms as in Eq.~(\ref{eq:direct-sum}); then 
the ladder operator ${\tilde{O}}_{\frac{3}{2},+}$ and the observable ${\tilde{O}}_{\frac{3}{2},3}$ should be expressed in terms of 
the linear combinations of ${\tilde{A}}_{\frac{3}{2},k}$ ($k=1, 4, 6$) and ${\tilde{\lambda}}_{\frac{3}{2},k}$ ($k=3,8,11$), respectively, as 
\begin{equation} 
	{\tilde{O}}_{\frac{3}{2},+} = \sum_{k=1,4,6}^6 c_k {\tilde{A}}_k, \label{eq:rise}
\end{equation} 
and 
\begin{equation} 
	{\tilde{O}}_{\frac{3}{2},3} = d_3 {\tilde{\lambda}}_{\frac{3}{2},3} + d_8 {\tilde{\lambda}}_{\frac{3}{2},8} + d_{11} {\tilde{\lambda}}_{\frac{3}{2},11}, \label{eq:z}
\end{equation} 
where $c_k$ and $d_k$ are the solutions of $[{\tilde{O}}_{\frac{3}{2},\pm}, {\tilde{O}}_{\frac{3}{2},3}]=\pm f{\tilde{O}}_{\frac{3}{2},\pm}$. 
The solutions, the number of subspaces $r$, the subspins $\{J_l \}$ in Eq.~(\ref{eq:direct-sum}), and the structure factor $f$ are respectively given by 
\begin{eqnarray}
	&\mathrm{(i)} \ & {\tilde{O}}_{\frac{3}{2},+} 
	= \sqrt{\frac{3}{10}} {\tilde{A}}_{\frac{3}{2},1} + \sqrt{\frac{2}{5}} {\tilde{A}}_{\frac{3}{2},4} + \sqrt{\frac{3}{10}} {\tilde{A}}_{\frac{3}{2},6}, \ 
	{\tilde{O}}_{\frac{3}{2},3} = {\tilde{\lambda}}_{\frac{3}{2},3}, \nonumber \\  & \ &r = 1, \ \{ J_1 = \frac{3}{2} \}, \ f = 1 \label{eq:type1} \\
	&\mathrm{(ii)} \ & {\tilde{O}}_{\frac{3}{2},+} = \frac{1}{\sqrt{2}} ({\tilde{A}}_{\frac{3}{2},1} \pm {\tilde{A}}_{\frac{3}{2},4}), \\ \nonumber & \ &
	{\tilde{O}}_{\frac{3}{2},3} = \frac{2}{\sqrt{10}} {\tilde{\lambda}}_{\frac{3}{2},3} + \frac{1}{\sqrt{2}} {\tilde{\lambda}}_{\frac{3}{2},8} 
	- \frac{1}{\sqrt{10}} {\tilde{\lambda}}_{\frac{3}{2},11}, \nonumber \\ & \ &r = 2, \ \{ J_1 = 1, J_2 = 0 \}, \ f = \sqrt{\frac{5}{2}}, 
	\label{eq:type2} \\
	&\mathrm{(iii)} \ & {\tilde{O}}_{\frac{3}{2},+} = \frac{1}{\sqrt{2}} ({\tilde{A}}_{\frac{3}{2},1} \pm {\tilde{A}}_{\frac{3}{2},6}), \ 
	{\tilde{O}}_{\frac{3}{2},3} = \frac{1}{\sqrt{5}} {\tilde{\lambda}}_{\frac{3}{2},3} + \frac{2}{\sqrt{5}} {\tilde{\lambda}}_{\frac{3}{2},11}, \nonumber \\ & \ &r = 2, \ 
	\{ J_1 = J_2 = \frac{1}{2} \}, \ f = \sqrt{5}, \label{eq:type3} \\
	&\mathrm{(iv)} \ & {\tilde{O}}_{\frac{3}{2},+} = {\tilde{A}}_{\frac{3}{2},1}, \ 
	{\tilde{O}}_{\frac{3}{2},3} = \frac{1}{\sqrt{10}} {\tilde{\lambda}}_{\frac{3}{2},3} + \frac{1}{\sqrt{2}} {\tilde{\lambda}}_{\frac{3}{2},8} 
	+ \sqrt{\frac{2}{5}} \ {\tilde{\lambda}}_{\frac{3}{2},11}, \nonumber \\ & \ &r = 3, \ \{ J_1 = \frac{1}{2}, J_2=J_3 = 0 \}, \ f =  \sqrt{10}. \label{eq:type4}
\end{eqnarray} 
The type (i) squeezing in Eq.~(\ref{eq:type1}) is equivalent to the spin squeezing among $\{ {\hat{J}}_x, {\hat{J}}_y, {\hat{J}}_z \}$ and the type (iii) squeezing in 
Eq.~(\ref{eq:type4}) is equivalent to the quadrupole-octupole squeezing among $\{{\hat{T}}_z^{\beta}, {\hat{T}}_{xyz}, \hat{Y} \}$ and the quadrupole squeezing 
among $\{{\hat{Q}}_{zx}, {\hat{Q}}_{yz}, \hat{Y} \}$. 

\subsection{Squeezing limits for four types of squeezing} 
In the case of the type (i) in Eq.~(\ref{eq:type1}), $r=1$ and the squeezing limit for the one-axis twisting is given by Eq.~(\ref{eq:minsqparam-r1}) as 
\begin{equation} 
	{\xi}_{\mathrm{min}}^2 \simeq \frac{1}{2} {\left ( \frac{1}{N} \right )}^{2/3} + \frac{1}{3N}, \label{eq:limit-type1}
\end{equation} 
which is achieved at the evolution time of ${\mu}_{\mathrm{min}}= \frac{2}{\sqrt{3}} \times N^{-2/3}$ corresponding to $t_{\mathrm{min}} = \frac{1}{\sqrt{3} \chi} \times N^{-2/3}$. 

\begin{figure}[t]
\hspace {0.2cm} (a) \hspace{6.8cm} (b) 
\begin{center}
\includegraphics[width=7.5cm,clip]{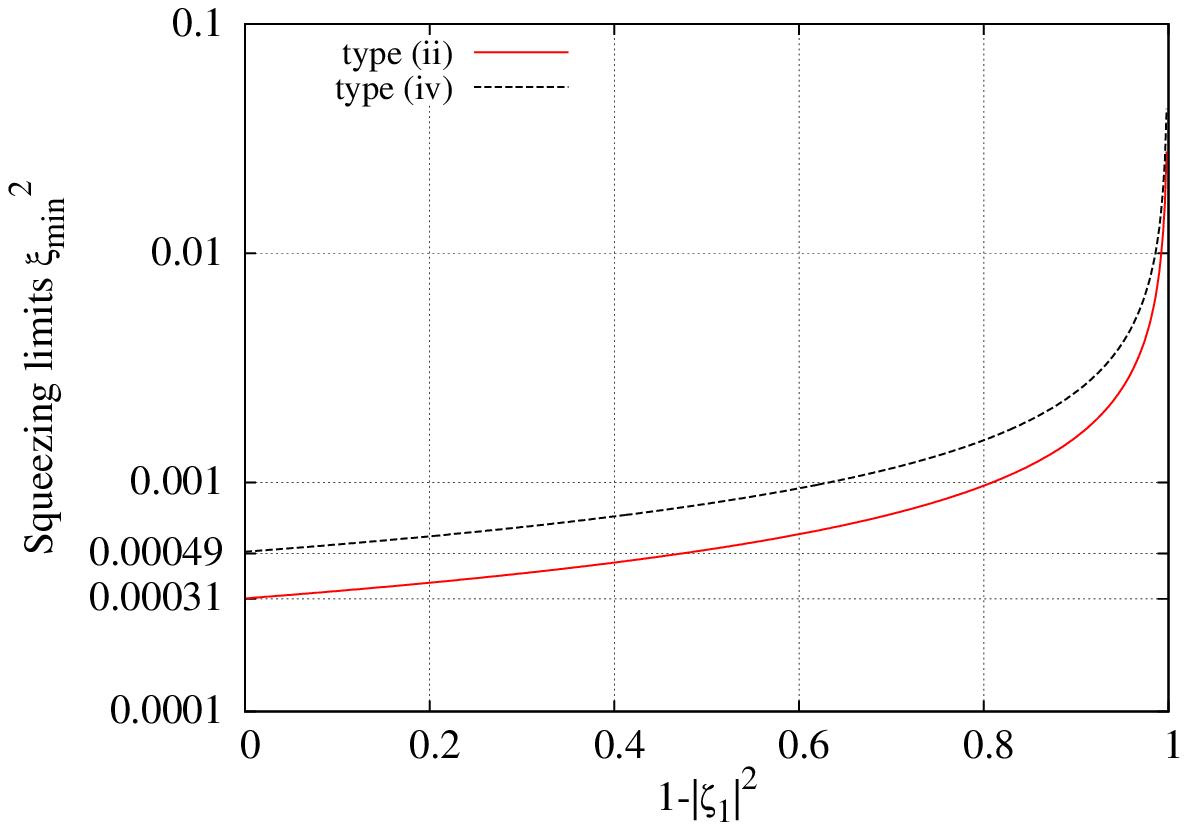}
\includegraphics[width=7.5cm,clip]{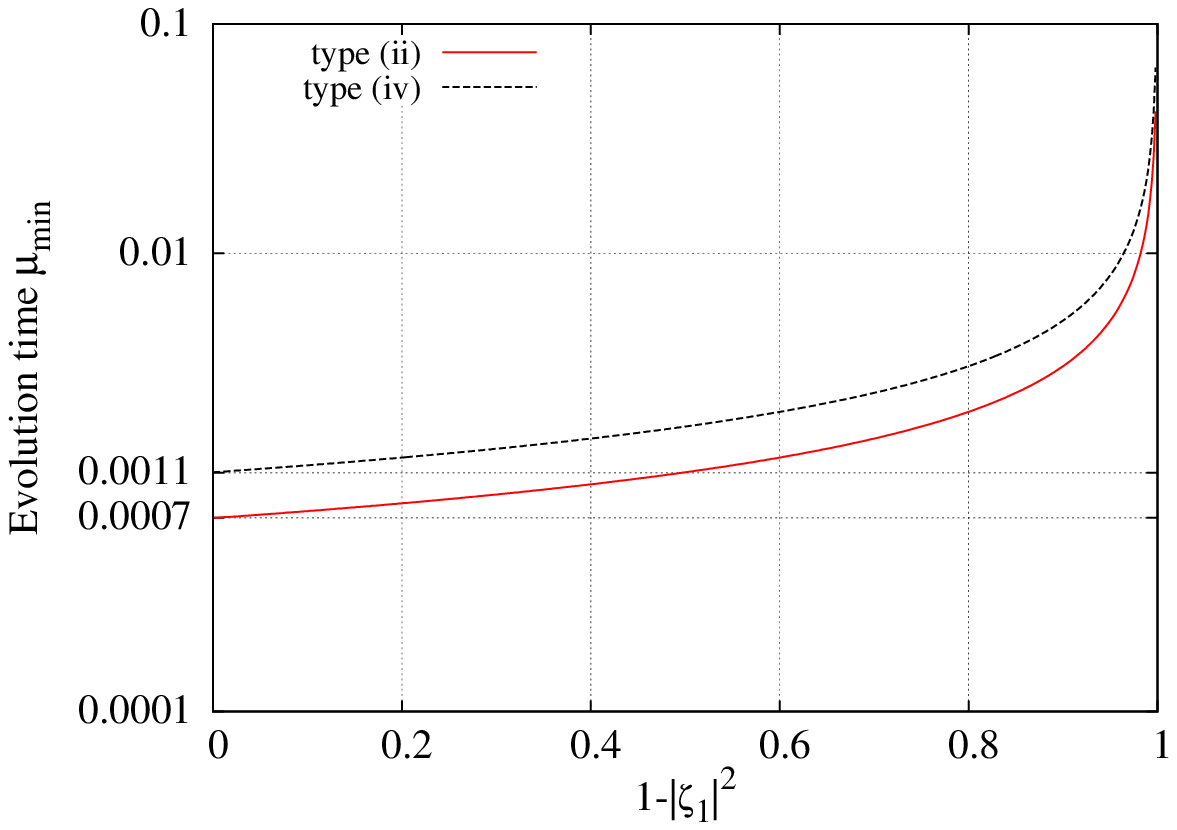}
\caption{(Color Online) (a) The $|{\zeta}_1|^2$-dependence on the squeezing limit ${\xi}^2_{\mathrm{min}}$ and (b) the corresponding evolution time 
${\mu}_{\mathrm{min}}$ for $N=10^5$. The horizontal dotted lines at ${\xi}_{\mathrm{min}}^2 = 0.00031$ and $0.00049$ in (a) and at ${\mu}_{\mathrm{min}} = 0.0007$ and 
$0.0011$ in (b) respectively indicate the squeezing limits for the type (ii) in Eq.~(\ref{eq:limit-type2}) and the type (iii) in Eq.~(\ref{eq:limit-type4}) and the corresponding evolution 
times when $|{\zeta}_1|^2=1$. }
\label{fig:sq(ii)and(iv)}
\end{center}
\end{figure} 
In the case of the types (ii)-(iv) in Eq.~(\ref{eq:type2})-(\ref{eq:type4}), the squeezing limits depend on the initial coherent state in Eq.~(\ref{eq:CSS-J}) in general; however, 
the squeezing limits for the types (ii) in Eq.~(\ref{eq:type2}) and (iv) in Eq.~(\ref{eq:type4}) can 
be calculated in the same manner as the type (i), when $|{\zeta}_l|^2 = {\delta}_{l1}$ in the initial state in Eq.~(\ref{eq:CSS-J}). 
They are given by 
\begin{eqnarray}
	&\mathrm{(ii)}  \ & {\xi}_{\mathrm{min}}^2 \simeq \frac{1}{2} {\left ( \frac{3}{2N} \right )}^{2/3} + \frac{1}{2N}, \label{eq:limit-type2} \\
	&\mathrm{(iv)} \ & {\xi}_{\mathrm{min}}^2 \simeq \frac{1}{2} {\left ( \frac{3}{N} \right )}^{2/3} + \frac{1}{N}, \label{eq:limit-type4}
\end{eqnarray} 
respectively. 
The minimum squeezing limits in Eqs.~(\ref{eq:limit-type2}) and (\ref{eq:limit-type4}) are achieved at ${\mu}_{\mathrm{min}} = 12^{1/6} \times N^{-2/3}$ 
($t_{\mathrm{min}} = \frac{12^{1/6}}{5\chi} \times N^{-2/3}$) and ${\mu}_{\mathrm{min}} = 2 \times 3^{1/6} \times N^{-2/3}$ ($t_{\mathrm{min}} = \frac{3^{1/6}}{10\chi} \times N^{-2/3}$), 
respectively. 
If $|{\zeta}_l|^2 \neq 0$ for $\exists l>0$, the squeezing limits for types (ii) and (iv) cannot be obtained by the expression in Eq.~(\ref{eq:minsqparam-r1}). 
We numerically calculate the $|{\zeta}_1|^2$-dependences of the squeezing limits and their corresponding evolution times and illustrate them in Figs.~\ref{fig:sq(ii)and(iv)} 
(a) and (b). 
In Figs.~\ref{fig:sq(ii)and(iv)}, we plot the squeezing limit ${\xi}_{\mathrm{min}}^2$ and the 
evolution time ${\mu}_{\mathrm{min}}$ with respect to $1-|{\zeta}_1|^2$. 
The squeezing limits for the types (ii) and (iv) monotonically decrease with increasing $|{\zeta}_1|^2$. 
For $|{\zeta}_1|^2 \simeq 1$, the squeezing limits are almost equal to Eqs.~(\ref{eq:limit-type2}) and (\ref{eq:limit-type4}), respectively; however, 
for $|{\zeta}_1|^2 < 0.2$, the minimum squeezing limits sharply increase due to the decreases 
in the number of the Schwinger bosons which are nonlinearly interacting via the one-axis twisting interactions in Eq.~(\ref{eq:OAT}). 

\begin{figure}[t]
\hspace {0.2cm} (a) \hspace{6.8cm} (b) 
\begin{center}
\includegraphics[width=7.5cm,clip]{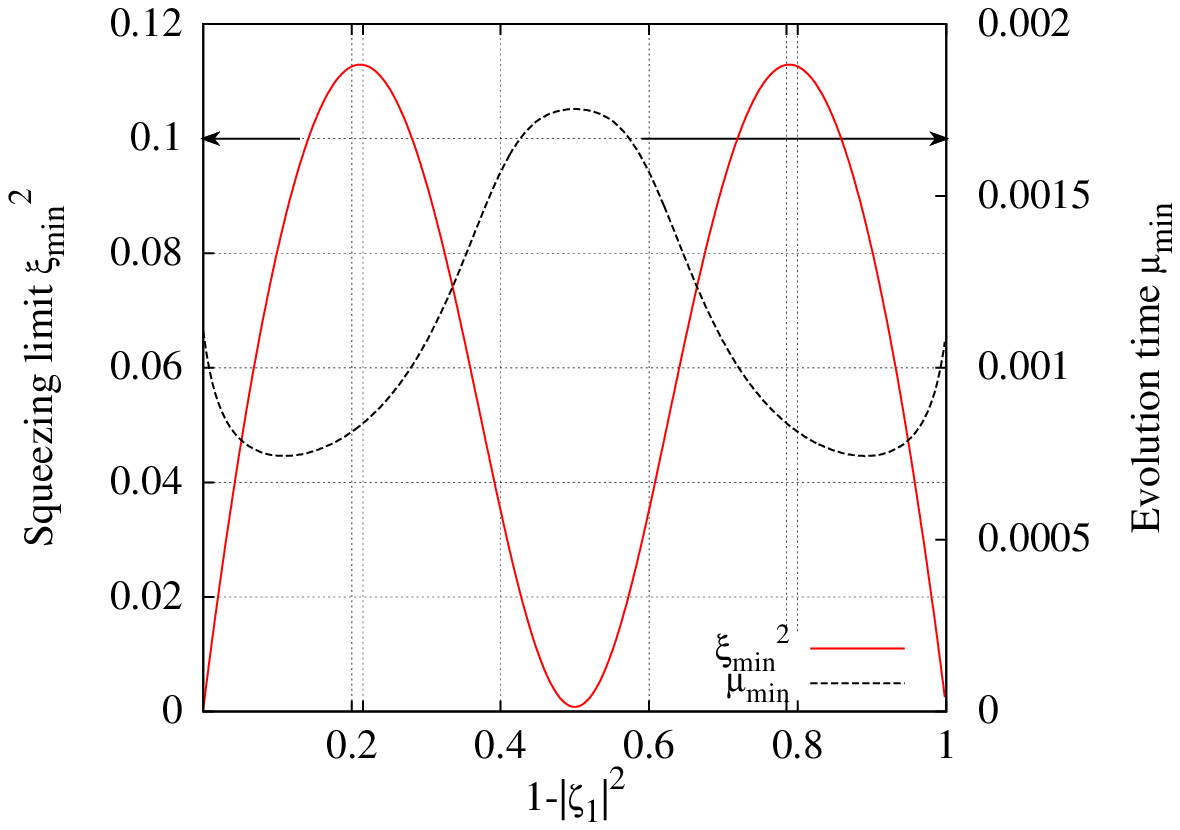}
\includegraphics[width=7.5cm,clip]{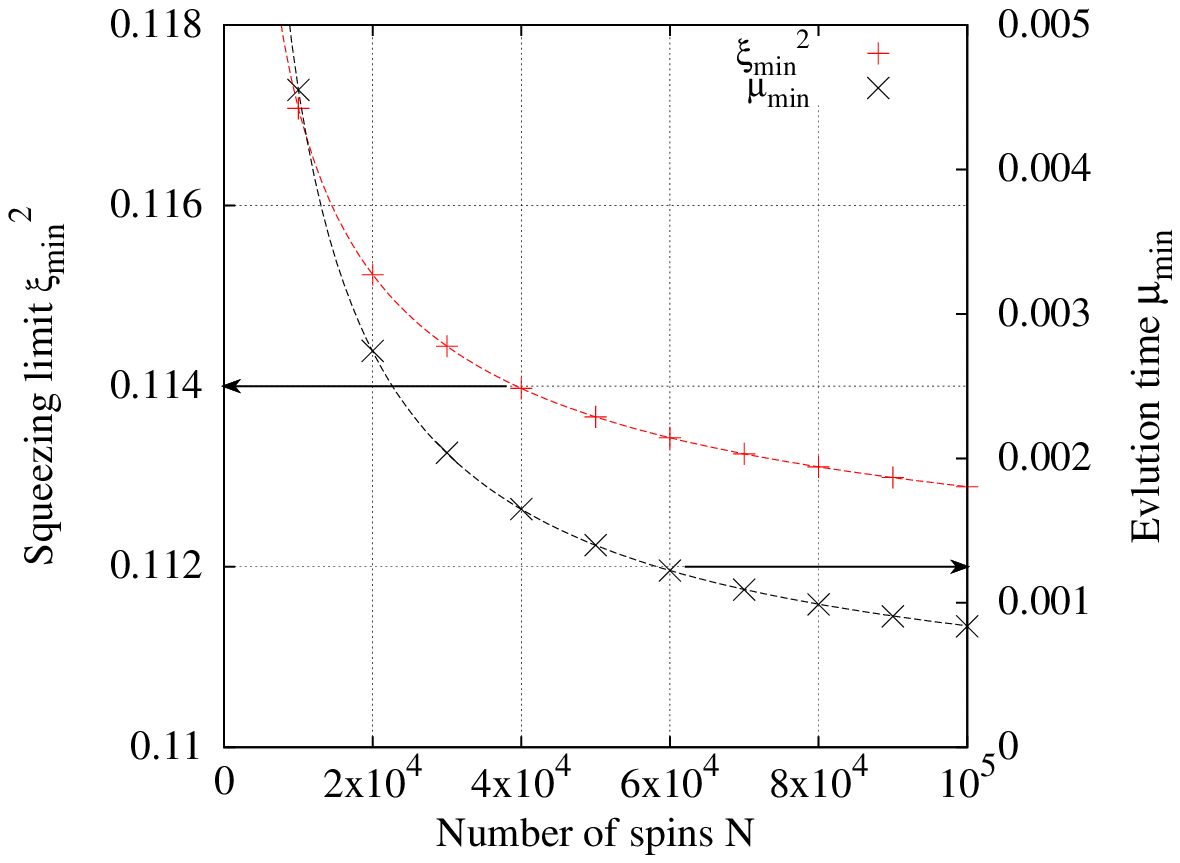}
\caption{(Color Online) (a) The $|{\zeta}_1|^2$-dependence of the squeezing limit ${\xi}^2_{\mathrm{min}}$ and the corresponding evolution time 
${\mu}_{\mathrm{min}}$ for $N=10^5$. The maxima of ${\xi}^2_{\mathrm{min}}$ 
are achieved at ${|{\zeta}_1|}^2 = 1 - \frac{\pi}{4}$ and $\frac{\pi}{4}$. (b) The $N$-dependence of the 
squeezing limit ${\xi}^2_{\mathrm{min}}$ and the corresponding evolution time ${\mu}_{\mathrm{min}}$ for ${|{\zeta}_1|}^2 = 1 - \frac{\pi}{4}$. 
The fitting function for ${\xi}^2_{\mathrm{min}}$ and ${\mu}_{\mathrm{min}}$ are given by Eq.~(\ref{eq:approx-xi}) and Eq.~(\ref{eq:approx-t}), respectively. }
\label{fig:sq(iii)}
\end{center}
\end{figure}

In the case of the type (iii) in Eq.~(\ref{eq:type3}), $r=2$ and $J_1 = J_2 = 1/2$, the $|{\zeta}_1|^2$-dependence of the minimum squeezing limit is periodic because of the symmetry 
with respect to the two subspaces. 
To see this, let us derive the expression for the squeezing limit for the type (iii): 
\begin{equation} 
	{\xi}^2 (\mu ) = \frac{1 + \frac{1}{4}  (N-1) \sum_{l=1}^2 {\Delta}_l(\mu )}{
	 \sum_{l=1}^2 {|{\zeta}_l|}^2 {(1 - 2 {|{\zeta}_l|}^2 {\sin}^2 \frac{\mu}{4})}^{N-1}} , \label{eq:sq-iii}
\end{equation} 
where ${\Delta}_l$'s ($l=1,2$) are defined as 
\begin{eqnarray} 
	& {\Delta}_l(\mu ) = \left [ 1 - {\left (1 - 2 {|{\zeta}_l|}^2 {\sin}^2 \frac{\mu}{2} \right )}^{N-2} \right ] \nonumber \\
	& \times \left \{ 1 - \sqrt{1 + {\left [ \frac{4 {|{\zeta}_l|}^2 \sin {\frac{\mu}{2}} \ {(1 - 2 {|{\zeta}_l|}^2 {\sin}^2 \frac{\mu}{4})}^{N-2}}{
	1 - {(1 - 2 {|{\zeta}_l|}^2 {\sin}^2 \frac{\mu}{2})}^{N-2} } \right ]}^2} \right \}. \label{eq:f-iii}
\end{eqnarray} 
When the initial state is given by $|{\zeta}_l|^2 = {\delta}_{l1}$, the squeezing limit is given by Eq.~(\ref{eq:limit-type4}) at 
${\mu}_{\mathrm{min}} =  2 \times 3^{1/6} \times N^{-2/3}$, which are same as those for the type (iv) with the initial state of $|{\zeta}_l|^2 = {\delta}_{l1}$, 
while the evolution time $t_{\mathrm{min}} = \frac{3^{1/6}}{5\chi} \times N^{-2/3}$ is two times larger than that for the type (iv) with the initial state of $|{\zeta}_l|^2 = {\delta}_{l1}$. 
When the initial state is the equal superposition of the two subspaces, i.e., ${|{\zeta}_l|}^2 = \frac{1}{2}$, the squeezing limit can be obtained by 
assuming $\alpha \gg 1$ and $\beta \ll 1$ to be 
\begin{equation} 
	{\xi}_{\mathrm{min}}^2 \simeq \frac{1}{2} {\left ( \frac{6}{N} \right )}^{2/3} + \frac{3}{N} \simeq \frac{1}{2} {\left ( \frac{6}{N} \right )}^{2/3}, \label{eq:limit-type3}
\end{equation} 
at the evolution time of ${\mu}_{\mathrm{min}} = 2 \times 3^{1/6} \times (N/2)^{-2/3}$ ($t_{\mathrm{min}} = \frac{48^{1/6}}{5\chi} \times N^{-2/3}$). 
Equation~(\ref{eq:limit-type3}) is $6^{2/3} \simeq 3.3$ times larger than the type (i) in Eq.~(\ref{eq:limit-type1}), $4^{2/3} \simeq 2.5$ times larger than the type (ii) in 
Eq.~(\ref{eq:limit-type2}) 
with the initial state of $|{\zeta}_l|^2 = {\delta}_{l1}$, and $2^{2/3} \simeq 1.6$ times larger than the type (iv) in Eq.~(\ref{eq:limit-type4}) 
with the initial state of $|{\zeta}_l|^2 = {\delta}_{l1}$ and the type (iii) with the initial 
state of $|{\zeta}_l|^2 = {\delta}_{l1}$. 
The $|{\zeta}_1|^2$-dependence of the squeezing limit and the corresponding evolution time for $N=10^5$ are illustrated in Fig.~\ref{fig:sq(iii)} (a). 
The squeezing limit reaches the maximum at $|{\zeta}_1|^2 \simeq 1-\frac{\pi}{4}$ and $\frac{\pi}{4}$. 
The dependence of the squeezing limit on the number of spins for $|{\zeta}_1|^2 \simeq 1-\frac{\pi}{4}$ is shown in Fig.~\ref{fig:sq(iii)} (b), which can 
be well fitted to 
\begin{equation} 
	{\xi}^2_{\mathrm{min}} \simeq 0.11 \pm 0.00 + \frac{0.57 \pm 0.00}{N^{0.50 \pm 0.00}} + \frac{3.8 \pm 0.0}{N} \label{eq:approx-xi}
\end{equation} 
by the least squared method. 
Equation~(\ref{eq:approx-xi}) implies that the scaling of the squeezing limit with respect to $N$ is $0$ for ${|{\zeta}_1|}^2 = \frac{\pi}{4}$ and 
$1 - \frac{\pi}{4}$, although the squeezing limit is still below the standard quantum limit of ${\zeta}^2 = 1$. 
The evolution time corresponding to the squeezing limit for ${|{\zeta}_1|}^2 = 1 - \frac{\pi}{4}$ can be well fitted to 
\begin{equation} 
	{\mu}_{\mathrm{min}} \simeq (3.9 \pm 0.0) \times N^{-0.73 \pm 0.00}, \label{eq:approx-t}
\end{equation} 
with respect to the number of spins $N$ by the least squared method. 

\section{Conclusion} 
In this paper, we consider the collective su($2J+1$) systems and classify the squeezing among the spin and multipolar observables generating the su($2$) subalgebra of the 
su($2J+1$) algebra, based on the unitary equivalence class of the su($2J+1$)-dimensional representations of  the observables. 
The matrix representations of the observables and their representation spaces can be decomposed into the direct sums of the lower dimensional irreducible representations of the 
su($2$) generators in Eq.~(\ref{eq:direct-sum}). 
This implies that if two sets of observables belong to the same unitary equivalence class, they can be decomposed into the same matrix representation in Eq.~(\ref{eq:direct-sum}) 
whose bases can be transformed to each other via an SU($2J+1$) transformation; hence they are characterized by the same subspins $\{J_l \}$ in Eq.~(\ref{eq:direct-sum}) giving the 
structure factor $f$ in Eq.~(\ref{eq:f}). 
The unitary equivalence class of the su($2$) subalgebra in the su($2J+1$) algebra can be found by choosing vertices in the Dynkin diagram of the su($2J+1$) algebra as shown 
in Fig.~\ref{fig:dynkin-sun}. 

The squeezing limits are determined by the dimensionality of the unitary equivalence class of the observables and the initial CSS involved by the squeezing. 
Taking the one-axis-twisted SSS for example, we calculate the squeezing limit ${\xi}^2_{\mathrm{min}}$, which is given by the function in Eq.~(\ref{eq:sqparam}) 
in terms of the subspins $\{ J_l \}$ in the irreducible representations in Eq.~(\ref{eq:direct-sum}) and the coefficients $\{ |{\zeta}_l|^2 \}$ of the initial CSS in 
Eq.~(\ref{eq:CSS-J}). 
When $|{\zeta}_l|^2 = {\delta}_{l1}$ in Eq.~(\ref{eq:CSS-J}), the squeezing limit ${\xi}^2_{\mathrm{min}}$ in Eq.~(\ref{eq:minsqparam-r1}) for the one-axis twisted 
SSS achieved to be proportional to $(J_{l_0}N)^{-2/3}$ at the evolution time of $\mu \equiv 2 \chi f^2 t \propto (J_{l_0}N)^{-2/3}$ in the limit of 
$J_{l_0}N \chi f^2 t \gg 1$ and $J_{l_0}N (\chi f^2 t)^2 \ll 1$, which implies that the squeezing among the observables, of which matrix representations are irreducible, gives the 
minimum squeezing limit of the collective su($2J+1$) consisting of $N$ spin-$J$ particles. 
In the case of $|{\zeta}_{l_0}|^2 < 1$ and $^{\exists} |{\zeta}_{l\neq {l_0}}|^2 \neq 0$, the analytical expressions of the squeezing limits in Eq.~(\ref{eq:sqparam}) cannot be easily 
obtained due to the interference between the representation spaces in Eq.~(\ref{eq:direct-sum}). 

Finally, we apply our classification to the squeezing in the collective su($4$) systems and obtain the squeezing limits analytically or numerically. 
The squeezing can be classified into one of four unitary equivalence classes as shown in Fig.~\ref{fig:su4-roots}. 
Their squeezing limits depends on the coefficients $\{ |{\zeta}_l|^2 \}$ in the initial coherent states in Eq.~(\ref{eq:CSS-J}) as well as the subspins $\{J_l \}$, whose behaviors were 
numerically calculated as shown in Figs.\ref{fig:sq(ii)and(iv)} (a) and \ref{fig:sq(iii)} (a). 
Since the subspins and the initial coherent sate reflect the structure of the unitary equivalence class of the spin and multipolar observables; hence the unitary equivalence class 
of the observables can be considered as one of the systematical ways to classify and quantify the squeezing. 

\ack
E. Y. thanks Prof. Mark Everitt, Prof. Todd Tilma, Dr. Shane Dooley, Mr. Itsik Cohen, and Ms. Marvellous Onuma-Kalu for fruitful discussions. 
This work is supported by MEXT Grant-in-Aid for Scientific Research(S) No. 25220601. 

\appendix 
\section{\label{a:0}Schwinger-boson approach to calculate expectation values for Eqs.~(\ref{eq:CSS-J}) and (\ref{eq:SSS-J})} 
The expectation values for the initial CSS in Eq.~(\ref{eq:CSS-J}) and for the one-axis-twisted SSS in Eq.~(\ref{eq:SSS-J}) 
can be simplified by the Schwinger boson approach. 

The observables $\{ {\hat{O}}_{J,k} \}$ can be decomposed into Eq.~(\ref{eq:direct-sum}), which are matrix-represented by the direct sums of the 
spin matrices $\{ {\tilde{\lambda}}_{J_l,k} \}$ for the spins $J_l$. 
For each of $r$ subspaces, we can define the Schwinger boson operator ${\hat{a}}_{l\pm}$ (${\hat{a}}_{l\pm}^{\dagger}$) which annihilates 
(creates) a boson in a mode `$l\pm$.' 
The annihilation (creation) Schwinger-boson operators 
${\hat{a}}_{l\pm}$ (${\hat{a}}_{l\pm}^{\dagger}$) satisfy 
\begin{equation} 
	[ {\hat{a}}_{ls}, {\hat{a}}_{l^{\prime}s^{\prime}} ] = 0, \  
	[ {\hat{a}}_{ls }, {\hat{a}}_{l^{\prime}s^{\prime}}^{\dagger} ] = {\delta}_{ll^{\prime}} {\delta}_{ss^{\prime}} \ 
	(s,s^{\prime} = \pm) , 
\end{equation} 
since the $r$ subspaces $V(\{|J_l, m_l {\rangle}_l \})$ ($l=1,\cdots ,r$) are orthogonal to each other. 
The $l$-th symmetric state $|\theta , \phi {\rangle}_l^{\otimes n_l}$ in Eq.~(\ref{eq:CSS-J}) can be regarded as a CSS of the $2J_ln_l$ spin-$1/2$ 
Schwinger bosons of the mode $l$ whose azimuth and polar angles are given by $\theta$ and $\phi$, respectively: 
\begin{eqnarray}
	|\theta , \phi {\rangle}_l^{\otimes n_l} 
	& = \sum_{m=0}^{N_l} \sqrt{_{N_l}C_m} \ {\cos}^{N_l-m} \frac{\theta}{2} \ {\sin}^m \frac{\theta}{2} \ e^{-im\phi} \nonumber \\
	&\times |n_{l+} = N_l - m, n_{l-} = m {\rangle}_{\mathrm{Sb}}, 
	\label{eq:coh-Sb}
\end{eqnarray} 
where $N_l \equiv 2J_ln_l$ represents the number of the $l$-th Schwinger bosons, and $|n_{l+}, n_{l-} {\rangle}_{\mathrm{Sb}}$ is the symmetric state of the $n_{l+}$ Schwinger 
bosons in the `$l+$' state and the $n_{l-}$ Schwinger bosons in the `$l-$' state. 
The matrix representations ${\tilde{\lambda}}_{J_l,k}$ for the $l$-th subspace with $J_l \neq 0$ can be mapped to the collective spin operators ${\hat{\Lambda}}_{J_l,k}$: 
\begin{eqnarray} 
	&{\tilde{\lambda}}_{J_l,1} \to {\hat{\Lambda}}_{J_l,1} = \frac{1}{2} ({\hat{a}}_{l+}^{\dagger} {\hat{a}}_{l-} + {\hat{a}}_{l-}^{\dagger} {\hat{a}}_{l+} ), \label{eq:L1}\\ 
	&{\tilde{\lambda}}_{J_l,2} \to {\hat{\Lambda}}_{J_l,2} = \frac{i}{2} (- {\hat{a}}_{l+}^{\dagger} {\hat{a}}_{l-} + {\hat{a}}_{l-}^{\dagger} {\hat{a}}_{l+} ), \label{eq:L2} \\ 
	&{\tilde{\lambda}}_{J_l,3} \to {\hat{\Lambda}}_{J_l,3} = \frac{1}{2} ({\hat{a}}_{l+}^{\dagger} {\hat{a}}_{l+} - {\hat{a}}_{l-}^{\dagger} {\hat{a}}_{l-} ), \label{eq:L3} 
\end{eqnarray} 
with the constraint ${\hat{\Lambda}}_{J_l,1}^2 + {\hat{\Lambda}}_{J_l,2}^2 +{\hat{\Lambda}}_{J_l,3}^2 = J_ln_l (J_ln_l+1)$. 
For $J_l = 0$, we define ${\hat{\Lambda}}_{J_l,1} = {\hat{\Lambda}}_{J_l,2} = {\hat{\Lambda}}_{J_l,3} = 0$. 
The observables in Eqs.~(\ref{eq:O3}) and (\ref{eq:O1}) can be expressed in terms of the Schwinger-boson representations in Eqs.~(\ref{eq:L1})-(\ref{eq:L3}) as 
\begin{eqnarray} 
	& {\hat{O}}_{J,\perp} = f \bigoplus_{l=1}^r \Bigl [ {\hat{\Lambda}}_{J_l,1} \cos {\phi} \sin {\theta} + {\hat{\Lambda}}_{J_l,2} \sin {\phi} \sin {\theta} + {\hat{\Lambda}}_{J_l,3} \cos {\theta} \Bigr ], \\ 
	& {\hat{O}}_{J,\nu} = f \bigoplus_{l=1}^r \Bigl [ {\hat{\Lambda}}_{J_l,1} (\cos {\phi} \cos {\theta} \cos {\nu} - \cos {\phi} \sin {\nu}) \nonumber \\ 
	&+ {\hat{\Lambda}}_{J_l,2} (\sin {\phi} \cos {\theta} \cos {\nu} + \cos {\phi} \sin {\nu}) - {\hat{\Lambda}}_{J_l,3} \sin {\theta} \cos {\nu} \Bigr ]. 
\end{eqnarray}
Thus, the expectation values in Eqs.~(\ref{eq:O3}) and (\ref{eq:O1}) for the CSS of Eq.~(\ref{eq:CSS-J}) can be obtained as Eqs.~(\ref{eq:exp-CSS}) and 
(\ref{eq:fluct-CSS}), respectively. 

Next, let us simplify the expectation values for the one-axis-twisted SSSs in Eq.~(\ref{eq:SSS-J}) in a manner similar to the case of the CSS. 
The one-axis twisting in Eq.~(\ref{eq:OAT}) and $|\frac{\pi}{2} , 0 {\rangle}_l^{\otimes n_l}$ in the initial CSS can be respectively expressed as 
\begin{equation}
	{\hat{H}}_{\mathrm{OAT}} = \hbar \chi f^2 { \left [ \bigoplus_{l=1}^r {\hat{\Lambda}}_{J_l,3} \right ]}^2 
	=  \hbar \chi f^2 \bigoplus_{l=1}^r {\hat{\Lambda}}_{J_l,3}^2, \label{eq:OAT-Sb}
\end{equation} 
and 
\begin{equation}
	|\frac{\pi}{2} , 0 {\rangle}_l^{\otimes n_l} 
	= \frac{1}{2^{N_l/2}} \sum_{m=0}^{N_l} \sqrt{_{N_l}C_m} \ |n_{l+} = N_l - m, n_{l-} = m {\rangle}_{\mathrm{Sb}}. 
	\label{eq:kthcoh-Sb}
\end{equation} 
The $l$-th one-axis twisting interaction $\hbar \chi f^2 {\hat{\Lambda}}_{J_l,3}^2$ in Eq.~(\ref{eq:OAT-Sb}) squeezes the $l$-th CSS 
in Eq.~(\ref{eq:kthcoh-Sb}). 
The one-axis-twisted SSS of the $l$-th Schwinger bosons at $\mu$ is given by 
\begin{eqnarray}
	|{\psi}_{\mathrm{OAT}} (\frac{1}{2},N_l; \mu) {\rangle}_l &\equiv \frac{1}{2^{N_l/2}} 
	\sum_{m=0}^{N_l} \sqrt{_{N_l}C_m} \ e^{-im\phi} e^{-\frac{i}{8} ({\hat{a}}_{l+}^{\dagger} {\hat{a}}_{l+} - {\hat{a}}_{l-}^{\dagger} {\hat{a}}_{l-} )^2 \mu} \nonumber \\
	&\times |n_{l+} = N_l - m, n_{l-} = m {\rangle}_{\mathrm{Sb}}. \label{eq:lthSSS}
\end{eqnarray}  
Here, we note that for an observable ${\hat{X}}_{J_l}$, two SSSs $|{\psi}_{\mathrm{OAT}} (\frac{1}{2},2J_ln_l; \mu ) {\rangle}_l$ 
and $|{\psi}_{\mathrm{OAT}} (\frac{1}{2},2J_ln_l^{\prime}; \mu ) {\rangle}_l$ in the $l$-th subspace satisfy 
\begin{equation} 
	\langle {\psi}_{\mathrm{OAT}} (\frac{1}{2},2J_ln_l^{\prime} ; \mu ) | {\hat{X}}_{J_l} |{\psi}_{\mathrm{OAT}} (\frac{1}{2},2J_ln_l; \mu ) {\rangle}_l 
	\propto {\delta}_{n_ln_l^{\prime}}, \label{eq:note}
\end{equation} 
since the expectation value vanishes when the numbers of the Schwinger bosons in the two states are not equal, i.e., $n_l \neq n_l^{\prime}$. 
Then, the expectation value of ${\hat{O}}_{J,1}$ can be calculated to give Eq.~(\ref{eq:perp}). 
The one-axis twisting redistribute the quantum fluctuations in the $O_{J,2}$-$O_{J,3}$ plane as follows: 
\begin{equation} 
	{\hat{O}}_{J,\nu} = {\hat{O}}_{J,2} \cos {\nu} - {\hat{O}}_{J,3} \sin {\nu}
	= f \bigoplus_{l=1}^r ( {\hat{\Lambda}}_{J_l,2} \cos {\nu} - {\hat{\Lambda}}_{J_l,3} \sin {\nu} ). \label{eq:obs}
\end{equation}
The quantum fluctuation in ${\hat{O}}_{J,\nu}$ with respect to the state $| {\Psi}_{\mathrm{OAT}} (J,N;\mu ) {\rangle}_{\mathrm{tot}}$ in Eq.~(\ref{eq:SSS-J}) is obtained by 
\begin{eqnarray} 
	\langle (\Delta O_{J, \nu} )^2 \rangle (\mu ) &= \langle  {\Psi}_{\mathrm{OAT}} (J,N;\mu ) | {\hat{O}}_{J,\nu}^2 | {\Psi}_{\mathrm{OAT}} (J,N;\mu ) 
	{\rangle}_{\mathrm{tot}} \nonumber \\
	&- \langle  {\Psi}_{\mathrm{OAT}} (J,N;\mu ) | {\hat{O}}_{J,\nu} | {\Psi}_{\mathrm{OAT}} (J,N;\mu ) {\rangle}_{\mathrm{tot}}^2. \label{eq:fluct}
\end{eqnarray} 
Here, the first term of the right-hand side of Eq.~(\ref{eq:fluct}) is given by 
\begin{eqnarray} 
	&\langle  {\Psi}_{\mathrm{OAT}} (J,N;\mu ) | {\hat{O}}_{J,\nu}^2 | {\Psi}_{\mathrm{OAT}} (J,N;\mu ) {\rangle}_{\mathrm{tot}} \nonumber \\
	&= f^2 \sum_{n_1=0}^N \sum_{n_2=0}^{N-n_1} \cdots \sum_{n_{r-1}=0}^{N-n_1-\cdots - n_{r-2}} \ _NC_{n_1} \ _{N-n_1}C_{n_2} \
	\cdots \ _{N-n_1-\cdots - n_{r-2}}C_{n_{r-1}} \nonumber \\
	&\times |{\zeta}_1|^{2n_1} |{\zeta}_2|^{2n_2} \cdots |{\zeta}_{r-1}|^{2n_{r-1}} |{\zeta}_r|^{2(N-n_1-\cdots - n_{r-1})} \nonumber \\
	&\times \sum_{l:J_l\neq 0} \langle  {\psi}_{\mathrm{OAT}} (\frac{1}{2},2J_ln_l;\mu ) | ({\hat{\Lambda}}_{J,2}\cos \nu - {\hat{\Lambda}}_{J,3}\sin \nu)^2 
	| {\psi}_{\mathrm{OAT}} (\frac{1}{2},2J_ln_l;\mu ) {\rangle}_l \nonumber \\
	&= f^2 \sum_{l:J_l\neq 0} \sum_{n_l=0}^N \ _NC_{n_l} |{\zeta}_l|^{2n_l} (1-|{\zeta}_l|^2)^{N-n_l} \nonumber \\
	&\times \langle  {\psi}_{\mathrm{OAT}} (\frac{1}{2},2J_ln_l;\mu ) | ({\hat{\Lambda}}_{J,2}\cos \nu - {\hat{\Lambda}}_{J,3}\sin \nu)^2 
	| {\psi}_{\mathrm{OAT}} (\frac{1}{2},2J_ln_l;\mu ) {\rangle}_l, \label{eq:fluct-1}
\end{eqnarray} 
where the first equality is derived from Eq.~(\ref{eq:note}) and the second equality is obtained by the symmetry with respect to the subspace index, $l$. 
Similarly to Eq.~(\ref{eq:fluct-1}), the second term in Eq.~(\ref{eq:fluct-2}) can be calculated as 
\begin{eqnarray} 
	&\langle  {\Psi}_{\mathrm{OAT}} (J,N;\mu ) | {\hat{O}}_{J,\nu} | {\Psi}_{\mathrm{OAT}} (J,N;\mu ) {\rangle}_{\mathrm{tot}} \nonumber \\
	&= f \sum_{l:J_l\neq 0} \sum_{n_l=0}^N \ _NC_{n_l} |{\zeta}_l|^{2n_l} (1-|{\zeta}_l|^2)^{N-n_l} \nonumber \\
	&\times \langle  {\psi}_{\mathrm{OAT}} (\frac{1}{2},2J_ln_l;\mu ) | ({\hat{\Lambda}}_{J,2}\cos \nu - {\hat{\Lambda}}_{J,3}\sin \nu)
	| {\psi}_{\mathrm{OAT}} (\frac{1}{2},2J_ln_l;\mu ) {\rangle}_l \nonumber \\
	&= 0, \label{eq:fluct-2}
\end{eqnarray} 
since $\langle  {\psi}_{\mathrm{OAT}} (\frac{1}{2},2J_ln_l;\mu ) | {\hat{\Lambda}}_{J,k} | {\psi}_{\mathrm{OAT}} (\frac{1}{2},2J_ln_l;\mu ) {\rangle}_l = 0$ for $k=2,3$. 
Substituting Eq.~(\ref{eq:lthSSS}) into Eq.~(\ref{eq:fluct-1}), we can simplify $\langle (\Delta O_{J, \nu} )^2 \rangle (\mu )$ in Eq.~(\ref{eq:fluct}) as Eqs.~(\ref{eq:fluct-3})-(\ref{eq:fluct-4}). 

\section{\label{a:1}Matrix representations of a single spin-3/2 operators} 
The matrix representations of the spin-vector components ${\tilde{J}}_{\mu}$ in Eq.~(\ref{eq:single-j}), the five independent components of the quadrupolar tensor, 
${\tilde{Q}}_{\mu \nu}$, ${\tilde{D}}_{xy}$, and $\tilde{Y}$ in Eqs.~(\ref{eq:single-q})-(\ref{eq:single-y}), and the seven independent components of the octupolar tensor, 
${\tilde{T}}^{\alpha}_{\mu}$, ${\tilde{T}}^{\beta}_{\mu}$, and ${\tilde{T}}_{xyz}$ in Eqs.~(\ref{eq:single-ta})-(\ref{eq:single-txyz}), are given by 
\begin{equation} 
\eqalign{
	{\tilde{J}}_{x} = \frac{1}{2} \left ( \begin{array}{cccc} 0 & \sqrt{3} & 0 & 0 \\ \sqrt{3} & 0 & 2 & 0 \\ 
	0 & 2 & 0 & \sqrt{3} \\ 0 & 0 & \sqrt{3} & 0 \end{array} \right ), \\ 
	{\tilde{J}}_{y} = \frac{i}{2} \left ( \begin{array}{cccc} 0 & -\sqrt{3} & 0 & 0 \\ \sqrt{3} & 0 & -2 & 0 \\ 
	0 & 2 & 0 & -\sqrt{3} \\ 0 & 0 & \sqrt{3} & 0 \end{array} \right ), \ 
	{\tilde{J}}_{z} = \frac{1}{2} \left ( \begin{array}{cccc} 3 & 0 & 0 & 0 \\ 0 & 1 & 0 & 0 \\ 
	0 & 0 & -1 & 0 \\ 0 & 0 &0 & -3 \end{array} \right ), \
} \label{eq:single-mj}
\end{equation} 
\begin{equation}
\eqalign{
	{\tilde{Q}}_{xy} = \frac{i\sqrt{5}}{2} \left ( \begin{array}{cccc} 0 & 0 & -1 & 0 \\ 0 & 0 & 0 & -1 \\ 
	1 & 0 & 0 & 0 \\ 0 & 1 & 0 & 0 \end{array} \right ) \ 
	{\tilde{Q}}_{yz} = \frac{i\sqrt{5}}{2} \left ( \begin{array}{cccc} 0 & -1 & 0 & 0 \\ 1 & 0 & 0 & 0 \\ 
	0 & 0 & 0 & 1 \\ 0 & 0 & -1 & 0 \end{array} \right ) \\
	{\tilde{Q}}_{zx} = \frac{\sqrt{5}}{2} \left ( \begin{array}{cccc} 0 & 1 & 0 & 0 \\ 1 & 0 & 0 & 0 \\ 
	0 & 0 & 0 & -1 \\ 0 & 0 & -1 & 0 \end{array} \right ) \ 
	{\tilde{D}}_{xy} = \frac{\sqrt{5}}{2} \left ( \begin{array}{cccc} 0 & 0 & 1 & 0 \\ 0 & 0 & 0 & 1 \\ 
	1 & 0 & 0 & 0 \\ 0 & 1 & 0 & 0 \end{array} \right ) \\
	\tilde{Y} = \frac{\sqrt{5}}{2} \left ( \begin{array}{cccc} 1 & 0 & 0 & 0 \\ 0 & -1 & 0 & 0 \\ 
	0 & 0 & -1 & 0 \\ 0 & 0 & 0 & 1 \end{array} \right ) 
} \label{eq:single-mq}
\end{equation} 
and 
\begin{equation} 
\eqalign{
	{\tilde{T}}^{\alpha}_{x} = \frac{1}{4} \left ( \begin{array}{cccc} 0 & -\sqrt{3} & 0 & 5 \\ 
	-\sqrt{3} & 0 & 3 & 0 \\ 0 & 3 & 0 & -\sqrt{3} \\ 5 & 0 & -\sqrt{3} & 0 \end{array} \right ), \\ 
	{\tilde{T}}^{\alpha}_{y} = \frac{i}{4} \left ( \begin{array}{cccc} 0 & \sqrt{3} & 0 & 5 \\ 
	-\sqrt{3} & 0 & -3 & 0 \\ 0 & 3 & 0 & \sqrt{3} \\ -5 & 0 & -\sqrt{3} & 0 \end{array} \right ), \
	{\tilde{T}}^{\alpha}_{z} = \frac{1}{2} \left ( \begin{array}{cccc} 1 & 0 & 0 & 0 \\ 
	0 & -3 & 0 & 0 \\ 0 & 0 & 3 & 0 \\ 0 & 0 & 0 & -1 \end{array} \right ), \\
	{\tilde{T}}^{\beta}_{x} = \frac{\sqrt{5}}{4} \left ( \begin{array}{cccc} 0 & -1 & 0 & -\sqrt{3} \\ 
	-1 & 0 & \sqrt{3} & 0 \\ 0 & \sqrt{3} & 0 & -1 \\ -\sqrt{3} & 0 & -1 & 0 \end{array} \right ), \\
	{\tilde{T}}^{\beta}_{y} = \frac{i\sqrt{5}}{4} \left ( \begin{array}{cccc} 0 & -1 & 0 & \sqrt{3} \\ 
	1 & 0 & \sqrt{3} & 0 \\ 0 & -\sqrt{3} & 0 & -1 \\ -\sqrt{3} & 0 & 1 & 0 \end{array} \right ), \\
	{\tilde{T}}^{\beta}_{z} = \frac{\sqrt{5}}{2} \left ( \begin{array}{cccc} 0 &0 & 1 & 0 \\ 
	0 & 0 & 0 & -1 \\ 1 & 0 & 0 & 0 \\ 0 & -1 & 0 & 0 \end{array} \right ), \
	{\tilde{T}}_{xyz} := \frac{i\sqrt{5}}{2} \left ( \begin{array}{cccc} 0 & 0 & -1 & 0 \\ 
	0 & 0 & 0 & 1 \\ 1 & 0 & 0 & 0 \\ 0 & -1 & 0 & 0 \end{array} \right ). \
} \label{eq:single-mt}
\end{equation}  

\section{\label{a:2}Root diagram and simple roots of the su($4$) algebra} 
First, we chose ${\tilde{\lambda}}_{\frac{3}{2},3}$, ${\tilde{\lambda}}_{\frac{3}{2},8}$, and ${\tilde{\lambda}}_{\frac{3}{2},11}$ as the Cartan subalgebra and obtain their 
adjoint representations $(\mathrm{ad} [{\tilde{\lambda}}_{\frac{3}{2},k_{\mathrm{C}}}])_{mn} \equiv f_{k_{\mathrm{C}}m}^n$ 
($k_{\mathrm{C}}=3,8,11$ and $m,n\neq 3,8,11$), where the structure constant $f_{k_{\mathrm{C}}m}^n$ is defined by 
$[{\tilde{\lambda}}_{k_{\mathrm{C}}}, {\tilde{\lambda}}_m] = i\sum_n f_{k_{\mathrm{C}}m}^n {\tilde{\lambda}}_n$. 
Here, the adjoint representations of ${\tilde{\lambda}}_{\frac{3}{2},k_{\mathrm{C}}}$ can be simultaneously diagonalized; hence they have the same eigenvectors 
${\tilde{A}}_{\frac{3}{2},k} = \sum_{k\neq 3,8,11} c_k {\tilde{\lambda}}_{\frac{3}{2},k}$ satisfying 
$[{\tilde{\lambda}}_{\frac{3}{2},k_{\mathrm{C}}}, {\tilde{A}}_{\frac{3}{2},k}] = {\mu}_{k_{\mathrm{C}}k} {\tilde{A}}_{\frac{3}{2},k}$, where 
${\mu}_{k_{\mathrm{C}}k}$ are the eigenvalues of 
$\mathrm{ad} [{\tilde{\lambda}}_{\frac{3}{2},k_{\mathrm{C}}}]$ corresponding to the eigenvectors ${\tilde{A}}_{\frac{3}{2},k}$. 
Then, we obtain twelve sets of eigenvalues ${\alpha}_k \equiv ({\mu}_{3k}, {\mu}_{8k}, {\mu}_{11k})$, i.e., the roots, and the eigenvectors 
${\tilde{A}}_{\frac{3}{2},k}$ ($k=1,\cdots 12$) corresponding to the roots. 
Plotting these roots in the Cartesian coordinate, we obtain the root diagram of the su($4$) algebra in Fig.~\ref{fig:su4-roots} (a). 
Here, the roots and their corresponding operators are given by 
\begin{eqnarray*} 
	{\balpha}_1 = \left ( \begin{array}{c} 1 \\ \sqrt{5} \\ 2 \end{array} \right ), \ {\tilde{A}}_{\frac{3}{2},1} 
	= \frac{\sqrt{15}}{10} {\tilde{J}}_+ + \frac{1}{2} {\tilde{Q}}_+ - \frac{\sqrt{15}}{20} {\tilde{T}}^{\alpha}_+ - \frac{1}{4} {\tilde{T}}^{\beta}_- 
	= \sqrt{5} E_{12}, \\ 
	{\balpha}_2 = \left ( \begin{array}{c} 2 \\ \sqrt{5} \\ -1 \end{array} \right ), \ {\tilde{A}}_{\frac{3}{2},2} 
	= \frac{1}{2} {\tilde{D}}_+ + \frac{1}{2} {\tilde{F}}_+ = \sqrt{5} E_{13}, \\
	{\balpha}_3 = \left ( \begin{array}{c} 3 \\ 0 \\ 1 \end{array} \right ), \ {\tilde{A}}_{\frac{3}{2},3} 
	= \frac{\sqrt{5}}{4} {\tilde{T}}^{\alpha}_- - \frac{\sqrt{3}}{4} {\tilde{T}}^{\beta}_+ = \sqrt{5} E_{14}, \\
	{\balpha}_4 = \left ( \begin{array}{c} 1 \\ 0 \\ -3 \end{array} \right ), \ {\tilde{A}}_{\frac{3}{2},4} 
	= \frac{1}{\sqrt{5}} {\tilde{J}}_+ + \frac{3}{4\sqrt{5}} {\tilde{T}}^{\alpha}_+ + \frac{\sqrt{3}}{4} {\tilde{T}}^{\beta}_- 
	= \sqrt{5} E_{23}, \\
	{\balpha}_5 = \left ( \begin{array}{c} 2 \\ -\sqrt{5} \\ -1 \end{array} \right ), \ {\tilde{A}}_{\frac{3}{2},5} 
	= \frac{1}{2} {\tilde{D}}_+ - \frac{1}{2} {\tilde{F}}_+ = \sqrt{5} E_{24}, \\
	{\balpha}_6 = \left ( \begin{array}{c} 1 \\ -\sqrt{5} \\ 2 \end{array} \right ), \ {\tilde{A}}_{\frac{3}{2},6} 
	= \frac{\sqrt{15}}{10} {\tilde{J}}_+ - \frac{1}{2} {\tilde{Q}}_- - \frac{\sqrt{15}}{20} {\tilde{T}}^{\alpha}_+ - \frac{1}{4} {\tilde{T}}^{\beta}_- 
	= \sqrt{5} E_{34}, \\
	{\balpha}_{6+k} = -{\balpha}_k, \ {\tilde{A}}_{6+k} = {\tilde{A}}_k^{\dagger}, \ (k=1,\cdots 6),   
\end{eqnarray*} 
where $E_{mn}$ denotes the matrix with 1 in the $mn$ entry and $0$s elsewhere and the ladder operators are defined by 
${\tilde{J}}_{\pm} \equiv {\tilde{J}}_x \pm i{\tilde{J}}_y$, 
${\tilde{Q}}_{\pm} \equiv {\tilde{Q}}_{zx} \pm i{\tilde{Q}}_{yz}$, 
${\tilde{D}}_{\pm} = {\tilde{D}}_{xy} \pm i {\tilde{Q}}_{xy}$, 
${\tilde{T}}^{\alpha}_{\pm} = {\tilde{T}}^{\alpha}_x \pm i{\tilde{T}}^{\alpha}_y$, 
${\tilde{T}}^{\beta}_{\pm} = {\tilde{T}}^{\beta}_x \pm i{\tilde{T}}^{\beta}_y$, 
and ${\tilde{F}}_{\pm} = {\tilde{T}}^{\beta}_z \pm i{\tilde{T}}_{xyz}$.

\end{document}